\documentclass[lineno]{jfm}

\usepackage{graphicx}
\usepackage{latexsym}
\usepackage[dvipsnames]{xcolor}
\usepackage{epstopdf, epsfig}
\usepackage{caption}
\usepackage{subcaption}
\usepackage{hyperref}
\usepackage{graphicx}
\usepackage{dcolumn}
\usepackage{bm}
\usepackage{amsmath}
\usepackage{amssymb}
\usepackage{placeins}
\usepackage{tikz}
\usepackage{pgfplots}
\usepackage{color}

\usepackage[abs]{overpic}
\usepackage{xcolor,varwidth}

\title{Rheology of concentrated suspension of fibers with load dependent friction coefficient}

\author{Monsurul Khan$^*$\aff{1},
  Rishabh V. More$^*$\aff{1},
  Arash Alizad Banaei\aff{2},
  Luca Brandt\aff{3}
  Arezoo M. Ardekani\aff{1},
   \corresp{\email{ardekani@purdue.edu}}
  }

\affiliation{\aff{1}School of Mechanical Engineering, Purdue University, West Lafayette, IN 47907, USA
\aff{2}Chemical Engineering, Stanford University, Stanford, CA 94305, USA
\aff{3} Flow and SeRC (Swedish e-Science Research Centre), Department of Engineering Mechanics, KTH, SE-100 44 Stockholm, Sweden
}

\begin{document}

\maketitle



             
\def\thefootnote{*}\footnotetext{Equal contributions in the writing of this manuscript}

\begin{abstract}
We investigate the effects of fiber aspect ratio, roughness, flexibility, and volume fraction on the rheology of concentrated suspensions in a steady shear flow using direct numerical simulations. We model the fibers as inextensible continuous flexible slender bodies with the Euler-Bernoulli beam equation governing their dynamics suspended in an incompressible Newtonian fluid. The fiber dynamics and fluid flow coupling is achieved using the immersed boundary method (IBM). In addition, the fiber surface roughness might lead to inter-fiber contacts resulting in normal and tangential forces between the fibers, which follow Coulomb's law of friction. The surface roughness is modeled as hemispherical protrusions on the fiber surfaces. Their deformation results in a normal load-dependent friction coefficient. Our simulations accurately predict the experimentally observed shear thinning in fiber suspensions. Furthermore, we find that the suspension viscosity $\eta$ \textcolor{black}{increases} with increasing the volume fraction, roughness, fiber rigidity, and aspect ratio. The increase in $\eta$ is the macroscopic manifestation of a similar increase in the microscopic contact contribution to the total stress with these parameters. In addition, we observe positive and negative first $N_1$ and second $N_2$ normal stress differences, respectively, with $|N_2| < |N_1|$, in agreement with previous experiments. Lastly, we propose a modified Maron-Pierce law to quantify the reduction in the jamming volume fraction by increasing the fiber aspect ratio and roughness. Our results and analysis establish the use of fiber surface tribology to tune the suspension flow behavior.
\end{abstract}

\maketitle


\section{Introduction}
Understanding the rheological properties of fiber suspensions is essential in many industrial applications such as paper and pulp production, biofuel production, and material reinforcement, to name a few \citep{bivins2005new,elgaddafi2012settling, hassanpour2012lightweight,lundell2011fluid,lindstrom2008simulation}.
These applications typically involve mixing, transportation, and handling concentrated fiber suspensions. Therefore, an accurate prediction of the suspension rheological properties is desirable to optimize these operations and prevent failures in industrial plants. However, fiber suspensions exhibit complex non-Newtonian behavior, such as the Weissenberg effect \citep{nawab1958viscosity, mewis1974rheological}, viscoelasticity \citep{thalen1964shear}, and shear thinning \citep{goto1986flow, kitano1981rheology, bounoua2016shear}, which makes their flow difficult to predict and control. A better understanding of the governing physical mechanisms and the effects of various controlling parameters on suspension rheology can help alleviate these difficulties. Over the years,
\textcolor{black}{different parameters such as volume fraction, fiber aspect ratio, and flexibility have been identified that directly influence the suspension rheology \citep{wu2010numerical,switzer2003rheology,banaei2020numerical,yamane1994numerical}}. Typically, the fibers are assumed to be perfectly smooth \citep{banaei2020numerical,wu2010numerical}. As a result, the details of inter-fiber interactions, especially surface roughness, have been ignored. However, even the so-called smoothest fibers have microscopic irregularities on their surfaces. These microscopic surface non-uniformities have been known to affect the bulk suspension rheology in dense sphere suspensions significantly \citep{tanner2016particle, hsu2018roughness,more2020roughness, more2020effect, more2020constitutive, hilali2022sheared}. However, their role in influencing fiber suspension rheology is not completely understood. 

Fiber suspensions are \textcolor{black}{classified as dilute, semi-dilute, or concentrated based on the number density defined as $\frac{nL^3}{V}$, where $\frac{n}{V}$ is the number of fibers per unit volume} and $L$ is the length of each fiber. In the dilute regime, i.e., $\frac{nL^3}{V}<1$, inter-fiber interactions are negligible. Interestingly, the stochasticity involved in the orientation changes of the fibers might result in a non-monotonic viscosity variation with volume fraction unlike in spherical particle suspensions \citep{blakeney1966viscosity}. In the semi-dilute regime, $1<\frac{nL^3}{V}<\frac{L}{d}$, where $d$ the fiber diameter, inter-fiber interactions start to influence the macroscopic properties and eventually become dominant in the concentrated regime $\frac{nL^3}{V}>\frac{L}{d}$. The viscosity of semi-dilute suspensions of rigid fibers in a Newtonian fluid is a function of the volume fraction but does not depend on the aspect ratio for sufficiently high shear rates \citep{bibbo1987rheology}. The experiments by \citet{djalili2006fibre} found a nearly constant viscosity as a function of the shear rate in the dilute regime, whereas shear thinning was observed in the semi-dilute regime. In addition to the shear rate-dependent viscosity, fiber suspensions also exhibit non-zero normal stress differences. In a shear flow, an isolated fiber can introduce a non-zero normal stress difference, the value of which is determined by the degree of instantaneous alignment of the fiber to the flow \citep{batchelor1971stress}. Parallel plate geometries have been used to measure the normal stress differences, $N_1-N_2$, where $N_1$ and $N_2$ are the first and second normal stress differences, respectively \citep{goto1986flow,petrich2000experimental,sepehr2004rheological}. These studies consistently reported a positive $N_1-N_2$, which increased with increasing particle volume fraction. Direct measurement of $N_1$ and $N_2$ observed an increase in their magnitudes when the aspect ratio decreases \citep{snook2014normal, bounoua2016normal}.

In the semi-dilute and concentrated suspensions, the contribution of non-hydrodynamic interactions, such as frictional contacts to the suspension stress, appears \citep{wu2010numerical}. Therefore, the
Jeffery’s model \citep{jeffery1922motion} to calculate the stress and viscosity for dilute suspensions considering only hydrodynamic interaction does not hold in semi-dilute and concentrated regimes \citep{anczurowski1967kinetics,okagawa1973kinetics}. Including contact interactions and the hydrodynamic interactions increases the viscosity and the first normal stress difference compared to the permissible range of Batchelor's theory \citep{batchelor1971stress}. Batchelor's theory \citep{batchelor1971stress} considered only hydrodynamic interactions to calculate the stress in concentrated elongated particle suspensions, and hence, under-predicts the increase in viscosity and the first normal stress difference \citep{salahuddin2013study}. \textcolor{black}{In suspensions, roughness might lead to contact and friction between the fibers, which lubrication interactions should prevent for perfectly smooth fibers.} Simulations of concentrated fiber suspensions with aspect ratio $11-32$ showed that the inter-particle forces are responsible for the existence of first normal stress differences \citep{butler2018microstructural, snook2014normal}. Moreover, increasing the inter-fiber friction coefficient between rigid fibers increases the fiber flocculation in the suspension, consequently increasing the suspension viscosity \citep{lindstrom2008simulation}. 
However, the simplified model of a constant friction coefficient employed in these previous studies does not capture the entire physics.  
\textcolor{black}{Experimental measurements have revealed that the friction coefficient is not constant but varies with the normal force and the roughness of the fibers \citep{brizmer2007elastic, huang2009measurement}. Hence, sophisticated models rooted in a physical understanding of the friction coefficient are needed to accurately model the inter-fiber contact dynamics. We utilize one such model in this study, as discussed in the later sections.} 

Fiber shape is another important factor governing suspension rheology. Slightly increasing the fiber curvature results in a change of the fiber rotation and a significant rise in the suspension viscosity compared to a straight fiber \citep{joung2001direct,goto1986flow}. These observations highlight the importance of fiber geometry and hence, fiber flexibility in determining the suspension rheological properties. Simulations to understand the role of flexibility by modeling fibers as a chain of rigid elements that can bend, stretch and twist observed a reduction in the period of fiber rotation with increasing flexibility \citep{yamamoto1993method}. On the bulk scale, the same model observed a decrease in the suspension viscosity with the fiber stiffness \citep{joung2001direct}. A decrease in the relative viscosity with the ratio of the shear rate to the stiffness of the fibers was reported using a rod chain model \citep{switzer2003rheology}. However, this model \cite{wu2010method, wu2010numerical} showed a decrease in the relative viscosity with fiber stiffness in contrast with the simulation by Switzer \textit{et al.} \cite{switzer2003rheology}. Thus, there is a lack of consistency and accuracy in the simulation results on the effect of fiber flexibility using the simple rod chain models. This suggests that more accurate modeling of fiber dynamics is required to capture the effects of various parameters, especially fiber flexibility. This can be achieved by modeling the fiber as a continuous object and is done in this study.

Transporting suspensions at high solid volume fractions is crucial to optimize processes and save energy in many industrial applications. However, long-lasting, dense, and system-spanning force chains can form at high solid volume fractions in frictional fiber flows, which strongly resist the flow. Jamming happens when the volume fraction is so high that the viscosity diverges and the suspension cannot flow irrespective of the applied high shear stress. Hence, it is difficult to characterize and predict the rheological properties of suspensions near jamming, which is apparent from the relatively small number of experimental studies for highly concentrated suspensions of fibers. In particular, experiments have shown that the increase in the aspect ratio of rigid fibers lowers the maximum volume fraction at which the suspension can flow \citep{tapia2017rheology}. They also observed that close to jamming volume fraction ($\phi_m$),  the viscosity scales as $(\phi_m-\phi)^{-1}$, where $\phi$ is the volume fraction of the suspension. \textcolor{black}{The dependence of the jamming on the fibers' surface roughness and flexibility was mainly addressed in dry suspensions \citep{guo2015computational, guo2020investigation} using the discrete element method, but hasn't been studied for wet suspensions, which are the focus of this study. Although understanding the suspension rheology at high volume fractions is of great industrial importance, the effects of aspect ratio and surface roughness on the jamming transition in suspensions of flexible fibers have not been quantified yet. We quantify these effects in the present study.} 

The suspension microstructure is influenced by the various governing factors mentioned above, and in turn, affects the bulk suspension rheology. Thus, understanding the relationship between these governing parameters, suspension microstructure, and macroscopic rheological properties of fiber suspensions is crucial to eliciting the underlying physical mechanisms governing the suspension rheology. Researchers have endeavored to understand the influence of some of these factors on the microstructure and rheological behavior of suspensions over the past decades, but we are still far from understanding the underlying physical mechanisms completely. To this end, we propose a numerical framework based on the immersed boundary method (IBM) \citep{peskin1972flow} and the discrete element method \citep{more2020effect} to simulate the shear flow of fiber suspensions and compute their flow properties. 
Our numerical model can predict the experimentally observed shear thinning behavior and quantify normal stress differences in fiber suspensions \citep{djalili2006fibre}. \textcolor{black}{This numerical framework enables us to perform a multidimensional parametric exploration to deconvolute and explicate the effects of fiber flexibility, aspect ratio, volume fraction, and roughness on the bulk suspension rheological properties.} The governing equations and the numerical model are explained in the next section, followed by results and a discussion of the effects of the governing parameters on the bulk rheological properties. 
\textcolor{black}{Finally, we elicit the functional forms of the dependence of the jamming volume fraction on the governing parameters, which can be useful in optimizing the transport and handling of dense fiber suspensions. }

\section{Governing equations and numerical model}\label{model}
In this section, we discuss the governing equations for the suspending fluid, fiber dynamics, and inter-fiber interactions. We use the method and algorithm in \citet{banaei2020numerical}, which we briefly discuss here. For detailed validation tests of the numerical method, the reader is referred to \citet{banaei2020numerical}.

\subsection{Flow field equations}
We consider an incompressible Newtonian suspending fluid, the flow of which is governed by the Navier-Stokes equations. In an inertial Cartesian frame of reference, the dimensionless momentum and mass conservation equations for an incompressible fluid are:

\begin{equation}
\frac{{\partial \mathbf{u}}}{{\partial t}} + \mathbf{\nabla} \cdot (\mathbf{u} \otimes \mathbf{u}) =  - \nabla p + \frac{1}{{{\mathop{\rm \textit{Re}}\nolimits} }}{\mathbf{\nabla} ^2}\mathbf{u} + \mathbf{f},
\label{NS}
\end{equation}
\begin{equation}
\mathbf{\nabla} \cdot \mathbf{u} = 0,
\end{equation}
where  $\mathbf u$ is the velocity field, $p$ the pressure, and $\mathbf f$ the volume force to account for the suspending fibers; $Re = \rho\dot{\gamma}L^2/\eta$ is the Reynolds number where $\rho$ and $\eta$ are the fluid density and dynamic viscosity, $L$ is the characteristic length scale which is also the fiber length, and $\dot{\gamma}$ is the applied shear rate. Details on  the fluid-structure interaction force $\mathbf{f}$ are given in Sec.~\ref{sec:nummethod}.

\subsection{Fiber dynamics}
We model the fiber as a continuous flexible slender body. The dynamics of a thin flexible fiber are described by the Euler-Bernoulli Beam equations under the constraint of inextensibility \citep{segel2007mathematics}. The equation of motion for each fiber in dimensionless form is:

\begin{eqnarray}
{\left( {\frac{{{\rho _{_f}}}}{{\rho{A_f}}}}\right)\frac{{{\partial ^2}\mathbf{X}}}{{\partial {t^2}}}} = &&
\frac{{{\partial ^2}\mathbf{X}_{fluid}}}{{\partial {t^2}}} +  \frac{\partial }{{\partial s}}(T\frac{{\partial \mathbf{X}}}{{\partial s}})- \frac{{{\partial ^2}}}{{\partial {s^2}}}(B\frac{{{\partial ^2}\mathbf{X}}}{{\partial {s^2}}})+\frac{{\Delta \rho }}{{\rho {A_f}}}Fr\frac{\mathbf{g}}{g} - \mathbf{F} + \mathbf{F}^f
\label{non-dimensional_fil}
\end{eqnarray}
where $s$ is the curvi-linear coordinate along the fiber, $\mathbf X = (x(s,t),y(s,t),z(s,t))$ is the position of the Lagrangian points on the fiber axis, $T$ the tension force along the fiber axis, $Fr=g/(L{\dot{\gamma}}^2)$ the Froude number with $\mathbf{g}$ the gravitational acceleration vector and $g = |\mathbf{g}|$, $B = \frac{EI}{\rho_{f}\dot\gamma^2L^4}$ the  dimensionless bending rigidity with $E$ the elastic modulus and $I$ the second moment of inertia for fiber cross-section, $\mathbf F$ the fluid-solid interaction force, $\mathbf {F}^f$  the net interaction force on the fiber due to neighboring fibers, $\rho_f$ the fiber linear density (mass per unit length), and $A_f$ the fiber cross sectional area. Equation \ref{non-dimensional_fil} is made dimensionless by the following characteristic scales: $L$ for length, $U_{\infty} = \dot{\gamma}L$ for velocity, $L/U_{\infty}$ for time, $\rho_f U_{\infty}^2$ for tension, $\rho_f U_{\infty}^2 L^2$ for bending and  $\rho_f U_{\infty}^2/L$ for force. \textcolor{black}{$\Delta\rho = {\rho _f} - \rho {A_f}$ denotes the linear density difference between the fibers and the surrounding fluid, with $\Delta\rho=0$ for the neutrally buoyant case. }
Finally, the inextensibity condition is expressed as
\begin{equation}
 \frac{{\partial \mathbf{X}}}{{\partial s}}.\frac{{\partial \mathbf X}}{{\partial s}} = 1.   
\label{inextensibility}
\end{equation}

\subsection{Numerical method}\label{sec:nummethod}
\begin{figure}
  \centerline{\includegraphics[width=0.7\linewidth]{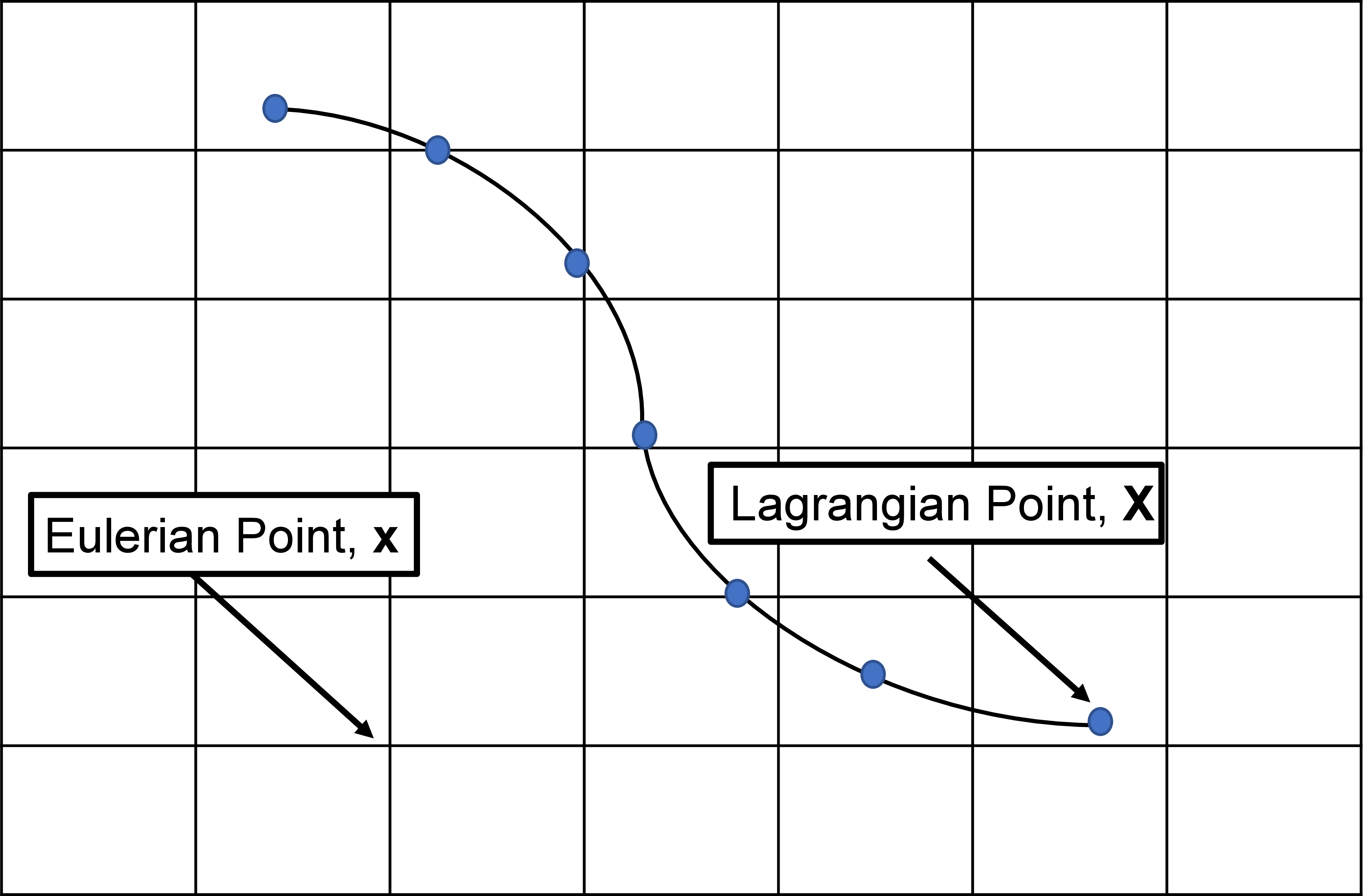}}
  \caption{Schematic of the Eulerian and the  Lagrangian grids. The black dots denote the Lagrangian points through which the position of the fibers are  defined.}
\label{fig:eulerian}
\end{figure}

The fluid-solid coupling is achieved using the Immersed Boundary Method (IBM) \citep{peskin1972flow}. In the IBM, the geometry of the object is represented by a volume force distribution $\mathbf{f}$ that mimics the effect of the object on the fluid. In this method, two sets of grid points are needed: a fixed Eulerian grid $\mathbf{x}$ for the fluid and a moving Lagrangian grid $\mathbf{X}$ for the flowing deformable structure as shown in figure \ref{fig:eulerian}. Each fiber has its own Lagrangian coordinate system. 
In this study we assume the fibers to be neutrally buoyant, so equation  (\ref{non-dimensional_fil}) is modified as: 
\begin{equation}
\frac{{{\partial ^2}\mathbf X}}{{\partial {t^2}}} = \frac{{{\partial ^2}{\mathbf X_{fluid}}}}{{\partial {t^2}}} + \frac{\partial }{{\partial s}}(T\frac{{\partial \mathbf X}}{{\partial s}}) - B\frac{{{\partial ^4}\mathbf X}}{{\partial {s^4}}} - \mathbf F + \mathbf {F}^f,
\label{pinal}
\end{equation}
where the LHS term is the acceleration of the fiber, and the RHS consists of the acceleration of the fluid particle at the fiber location and the different forces acting on the fibers. \textcolor{black}{This is done to avoid the singularity of the coefficient matrix in the fiber equation of motion (equation \ref{non-dimensional_fil}) for the neutrally buoyant case, as reported by  \citet{pinelli2017pelskin}.} To solve for the fiber position in equation \ref{pinal}, we first solve the Poisson's equation for tension \citep{huang2007simulation} obtained by combining equations \ref{inextensibility} and \ref{pinal}, which is given as:
\begin{eqnarray}
\frac{{\partial \mathbf X}}{{\partial s}}.\frac{{{\partial ^2}}}{{\partial {s^2}}}(T\frac{{\partial \mathbf X}}{{\partial s}}) =&&
\frac{1}{2}\frac{{{\partial ^2}}}{{\partial {t^2}}}(\frac{{\partial \mathbf X}}{{\partial s}}.\frac{{\partial \mathbf X}}{{\partial s}})- \frac{{{\partial ^2}\mathbf X}}{{\partial t\partial s}}.\frac{{{\partial ^2}\mathbf X}}{{\partial t\partial s}} - \frac{{\partial \mathbf X}}{{\partial s}}.\frac{\partial }{{\partial s}}(\mathbf{F}^a + \mathbf{F}^b + \mathbf{F}^f - \mathbf F),
\label{Poission}
\end{eqnarray}
where $\mathbf{F}^a = \frac{{{\partial ^2}\mathbf{X_{fluid}}}}{{\partial {t^2}}}$ is the acceleration of the fluid particle at the fiber location and $\mathbf{F}^b =  - B\frac{{{\partial ^4}\mathbf{X_{}}}}{{\partial {s^4}}}$ is the bending force. 
As the fibers are freely suspended in the fluid medium, we impose zero force, torque and tension at the free ends. So,
\begin{equation}
    \frac{{{\partial ^2}\mathbf X}}{{\partial {s^2}}} = 0, \frac{{{\partial ^3}\mathbf X}}{{\partial {s^3}}} = 0, \textrm{ and } T = 0.
\label{BC}    
\end{equation}
At each time step, the fluid velocity is first interpolated onto the Lagrangian grid points using the smooth Dirac delta function, $\delta$ \citep{roma1999adaptive}:
\begin{equation}
   {\mathbf U_{ib}} =  \int\limits_V {{\mathbf  u(\mathbf x,t)\delta (\mathbf X - \mathbf x)dV}}, 
\end{equation}
The fluid and solid equations are then coupled by the fluid-solid interaction force, 
\begin{equation}\label{eq:FSI_force}
    \mathbf F = \frac{{\mathbf U - {\mathbf U_{ib}}}}{{\Delta t}},
\end{equation}
where  $\mathbf U_{ib}$ is the interpolated fluid velocity at the Lagrangian points defining the fibers, $\mathbf U$ is the velocity of the Lagrangian points and $\Delta t$ is the time step. 
The Lagrangian force is then extrapolated onto the fluid grid as
\begin{equation}
    \mathbf f(\mathbf x,t) = \frac{\pi}{4}r_p^2\int\limits_L {\mathbf  F(\mathbf x,t)\delta (\mathbf X - \mathbf x)ds}.
\end{equation}
Here $r_p = d/L$, is the slenderness ratio of the fiber, which is the inverse of its aspect ration defined as $AR=L/d$. The interaction force can be further split into $\mathbf F^f=\mathbf F^{lc}+\mathbf F^c$, where $\mathbf F^{lc}$ and $\mathbf F^c$ are the lubrication correction, and contact interaction, respectively, as explained below.

\subsection{Lubrication interactions}
To accurately resolve the lubrication interactions between the fibers when the inter-fiber gap falls below a few grid-sizes, we use the lubrication correction model of \citet{lindstrom2008simulation}. \textcolor{black}{The lubrication force model is based on the solution for two infinitely long cylinders adapted to two different configurations: parallel cylinders or at an arbitrary angle}. The first-order approximation of the lubrication force for the non-parallel case was derived by \citet{yamane1994numerical} and is given as follow:
\begin{equation}
    \mathbf F_1^l = \frac{{ - 12}}{{{\mathop{\rm Re}\nolimits} \sin \alpha }}\frac{{\dot{\mathbf{h}}}}{h},
\label{lub1}    
\end{equation}
where $h$ is the shortest distance between the cylinders, $\dot{\mathbf{h}}$ is the relative normal velocity between the closest points on the fibers, and $\alpha$ is the contact angle. The first order approximation of the lubrication force per unit length between parallel cylinders was derived by \citet{kromkamp2005shear}:
\begin{equation}
    \begin{array}{l}
\mathbf F_2^l = \frac{{ - 4}}{{\pi {\mathop{\rm Re}\nolimits} r_p^2}}\left( {{A_1} + {A_2}\frac{h}{a}} \right){\left( {\frac{h}{a}} \right)^{ - 3/2}}\dot{\mathbf{h}},\\
{A_1} = 3\pi \sqrt 2/8, {A_2} = 207\pi \sqrt 2 /160,
\end{array}
\label{lub2}
\end{equation}
here $a$ is the cylinder radius $(a = d/2)$. Based on equations \ref{lub1} and \ref{lub2}, the following approximation of the lubrication force for two finite cylinders is \citep{lindstrom2008simulation}:
\begin{equation}
    {\mathbf F^l} = \min \left( {\mathbf F_1^l/\Delta s,\mathbf F_2^l} \right).
\end{equation}
The force for the non-parallel case is divided by the Lagrangian grid spacing $\Delta s$ to calculate the force per unit length.
The numerical method implemented to calculate the lubrication force is discussed in \citet{banaei2020numerical}, and hence is not repeated here. As the distance between the fibers becomes of the order of the mesh size, hydrodynamic interactions are not well resolved; to address this issue, we introduce a lubrication correction force \citep{lindstrom2008simulation}. When the shortest distance between two Lagrangian points becomes lower than $d/4$, we introduce lubrication correction as: $\mathbf F^{lc}=\mathbf F^{l}-\mathbf F_0^l$, where $F_0^l$ is the lubrication force at a $d/4$ distance.

The lubrication forces diverge as the minimum inter-fiber separation decreases, and theoretically should prevent the fibers from coming into direct contacts. However, the thin lubrication film between close fibers can break because of the presence of irregularities on their surfaces leading to a direct contact between the fibers and hence, contact forces. Direct contacts also give rise to non-negligible friction between the fibers. The following section provides an overview of the implemented contact model.

\subsection{Contact model}

\begin{figure*}
\centering
\begin{subfigure}{0.5\textwidth}
  \centering
  \includegraphics[width=1.0\linewidth]{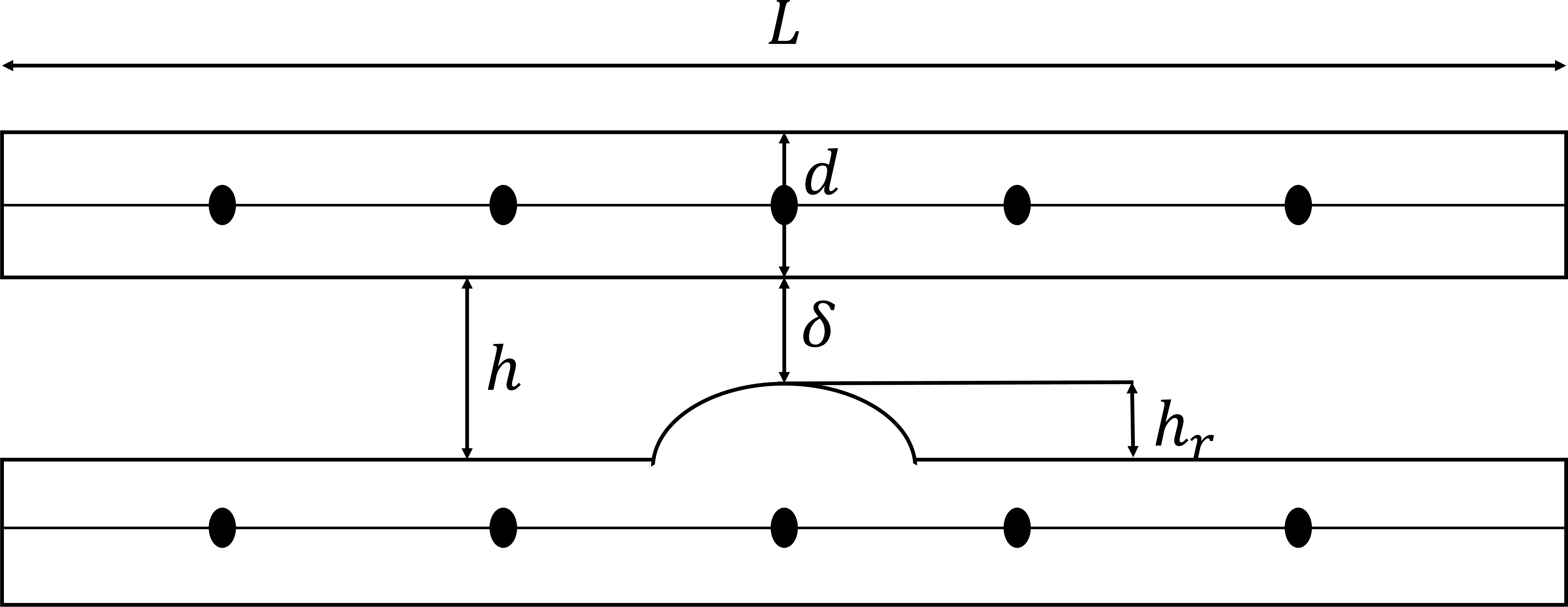}
  \caption{}
  \label{fig:deformation}
\end{subfigure}
~
\begin{subfigure}{0.5\textwidth}
  \centering
  \includegraphics[width=1.0\linewidth]{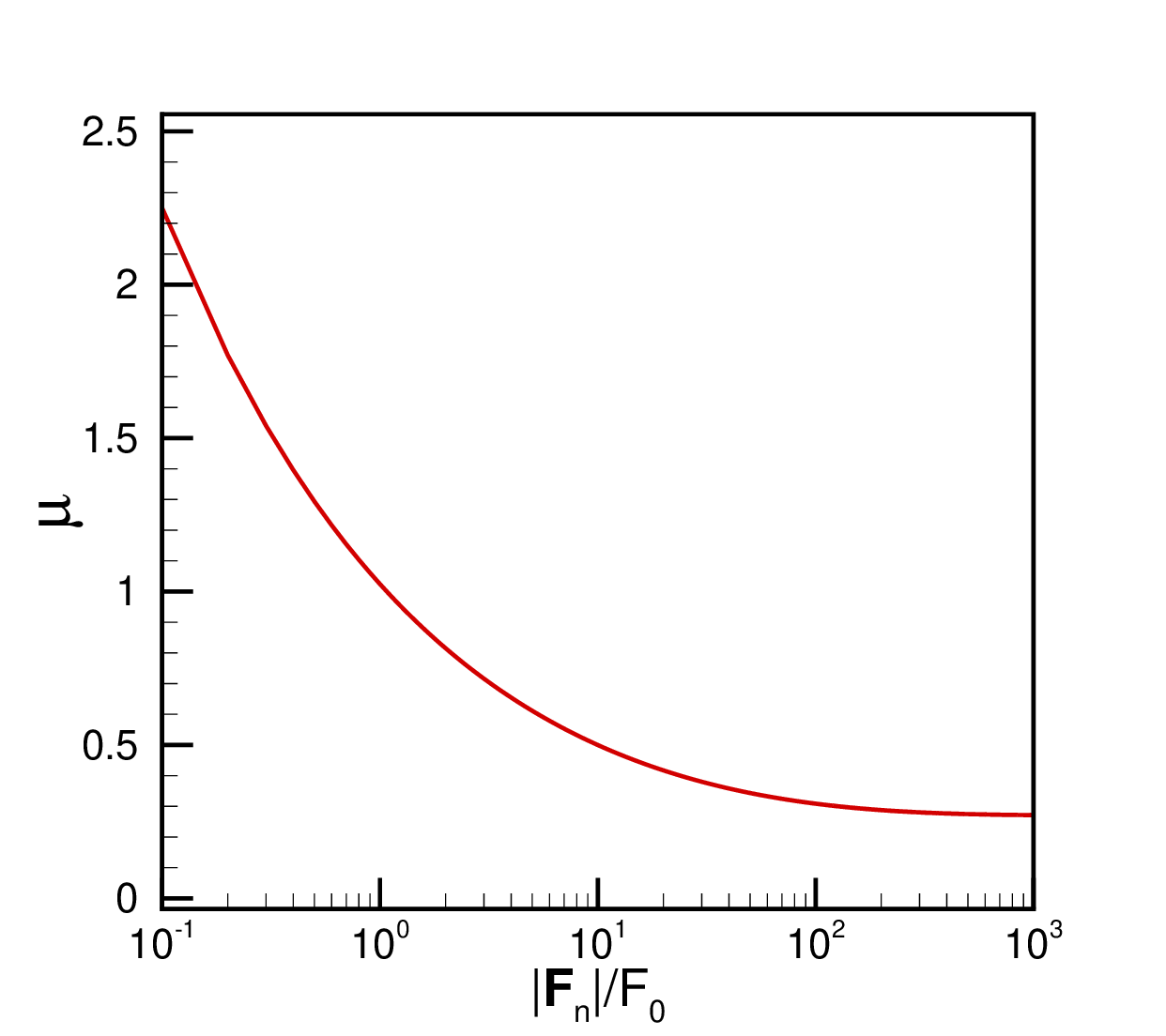}
  \caption{}
  \label{fig:mu_normal_force}
\end{subfigure}
\caption{(a) A sketch of the roughness model, $L$ and $d$ are the length and diameter of the fiber, respectively, $h_r$ is the roughness height, and $\delta=h-h_r$ is the surface overlap. Contact occurs when $\delta \le 0$. Dots along the axes of the fibers indicate Lagrangian points. (b) Friction coefficient $\mu$ as a function of the dimensionless contact normal force (equation \ref{eq:frictionlaw}). }
\label{fig:sharratebudget}
\end{figure*}

With increasing volume fraction, the surrounding fibers hinder the free rotation of nearby fibers, giving rise to fiber-fiber contacts that influences the micro-structure. Hence, the importance of accurately modelling the contact dynamics cannot be over-stressed, as routinely done when studying dry fiber suspensions \citep{guo2015computational} and more recently spherical particle suspensions \citep{gallier2014rheology, more2020effect}. In recent years, researchers have investigated the effects of inter-particle contacts on the rheology of suspensions by using experimentally validated models, which have been proven to be effective in quantitatively reproducing many experimental findings \citep{mari2014shear, more2020effect, more2020roughness}. Specifically, the single-asperity model of the surface roughness has been widely used owing to its simplicity and effectiveness \citep{gallier2014rheology, lobry2019shear, more2020constitutive, more2020roughness, more2020effect}. Hence, we take the same approach and model the asperity as a hemispherical bump on the fiber surface as shown in figure \ref{fig:deformation}. Actual asperities might not be just hemispherical and can come in a variety of geometries \citep{tanner2016particle}. However, on average we can model their behavior by approximately assuming them hemispherical as routinely done in the tribology literature \citep{broedersz2014modeling}. 

Let us consider two fibers as shown in figure~\ref{fig:deformation} with diameter $d$ and length $L$ having surface roughness height $h_r$.
The contact between the fibers occurs when the smallest inter-fiber separation distance $h$ becomes smaller than $h_r$. As a result, the asperity deforms which results in forces normal, $\mathbf F_n$, and tangential, $\mathbf F_t$ to the fiber surface. The normal contact force is given by the Hertz law:
\begin{equation}\label{eq:hertz}
\mathbf F_n=- F_0\left(\frac{|\delta|}{L}\right)^{3/2} \mathbf{n},
\end{equation}
where $\delta$ is the surface overlap defined as $\delta=h-h_r$, and $F_0/L^{3/2}$ is the normal stiffness, which is a function of the fiber material properties, and the roughness size \citep{lobry2019shear, more2020effect}. We use $F_0$ as the characteristic contact force scale in this study. The contact happens when $\delta \leq 0$. The Coulomb's friction law gives the tangential force:
\begin{equation}
\mathbf F_t=\mu|\mathbf F_n|\frac{\mathbf F_t}{|\mathbf F_t|},
\end{equation}
where $\mu$ is the friction coefficient. \textcolor{black}{Several studies have assumed} a constant friction coefficient when numerically studying the dynamics of fiber suspensions \citep{banaei2020numerical,stickel2009rheology}. 
In reality, the coefficient of friction depends on many factors such as the fiber material and the roughness size, which are implicitly included in the normal force via the normal stiffness $F_0/L^{3/2}$ \citep{lobry2019shear, more2020effect}. Hence, a normal load dependent coefficient of friction is a more accurate physical description than a simple constant. We use the Brizmer model \citep{brizmer2007elastic} for $\mu$, which was derived from the measurements between a hemisphere and a flat surface and validated using the finite element analysis \citep{brizmer2007elastic}, which makes it applicable to a wide range of materials and conditions \citep{brizmer2007elastic,lobry2019shear,more2020effect}. It is given by:

\begin{equation}
\mu  = 0.27\coth \left[ {{0.27{\left({\frac{{|\mathbf F_n^{(i,j)}|}}{{{ F_0}}}} \right)}^{0.35}}} \right],
\label{eq:frictionlaw}
\end{equation}
where $F_0$ is the characteristic contact force scale introduced in equation \ref{eq:hertz}. Equation~\ref{eq:frictionlaw} is a decreasing function of $|\mathbf F_n^{(i,j)}|/F_0$ as illustrated in figure~\ref{fig:mu_normal_force}. \textcolor{black}{Thus, the coefficient of friction decreases with increasing the normal force between the contacting fibers (which is equivalent to increasing asperity deformation in equation~\ref{eq:hertz}) and attains a plateau at high normal loads.} \textcolor{black}{ Moreover, to avoid overlap of the fibers with the walls, we implement a repulsive force having the form of a Morse potential as:  
\begin{equation}
    \emptyset  = {D_e}[{{e^{ - 2\beta ({r_f} - {r_e})}}} - 2{e^{ - \beta ({r_f} - {r_e})}},
\end{equation}
where $D_e$ is the interaction strength, $\beta$ a geometric scaling factor, $r_f$ the distance between a point on a fiber and the wall, and $r_e$ the zero-cut-off force distance. The repulsive between the elements i and j is the derivative of the potential function, 
\begin{equation}
    {F_r} =  - \frac{{d\emptyset }}{{dr}}{d_{ij}},
\end{equation}}

\textcolor{black}{where $d_{ij}$ is the unit vector in the direction joining the contact point. }

After calculating the short range interactions between fibers, the tension inside each fiber is computed by solving equation \ref{Poission}. Then, the new fiber position is obtained by solving equation \ref{pinal} and the fluid equations are advanced in time. The fluid equations are solved with a second-order finite difference method on a fix staggered grid. The equations are advanced in time by a semi-implicit fractional step-method, where the second order Adams-Bashforth method is used for the convective terms, a Helmholtz equation is built with the diffusive and temporal terms, and all other terms are treated explicitly \citep{alizad2017numerical}.

\subsection{Boundary conditions and Domain size}

\begin{figure}
  \centerline{\includegraphics[width=0.8\linewidth]{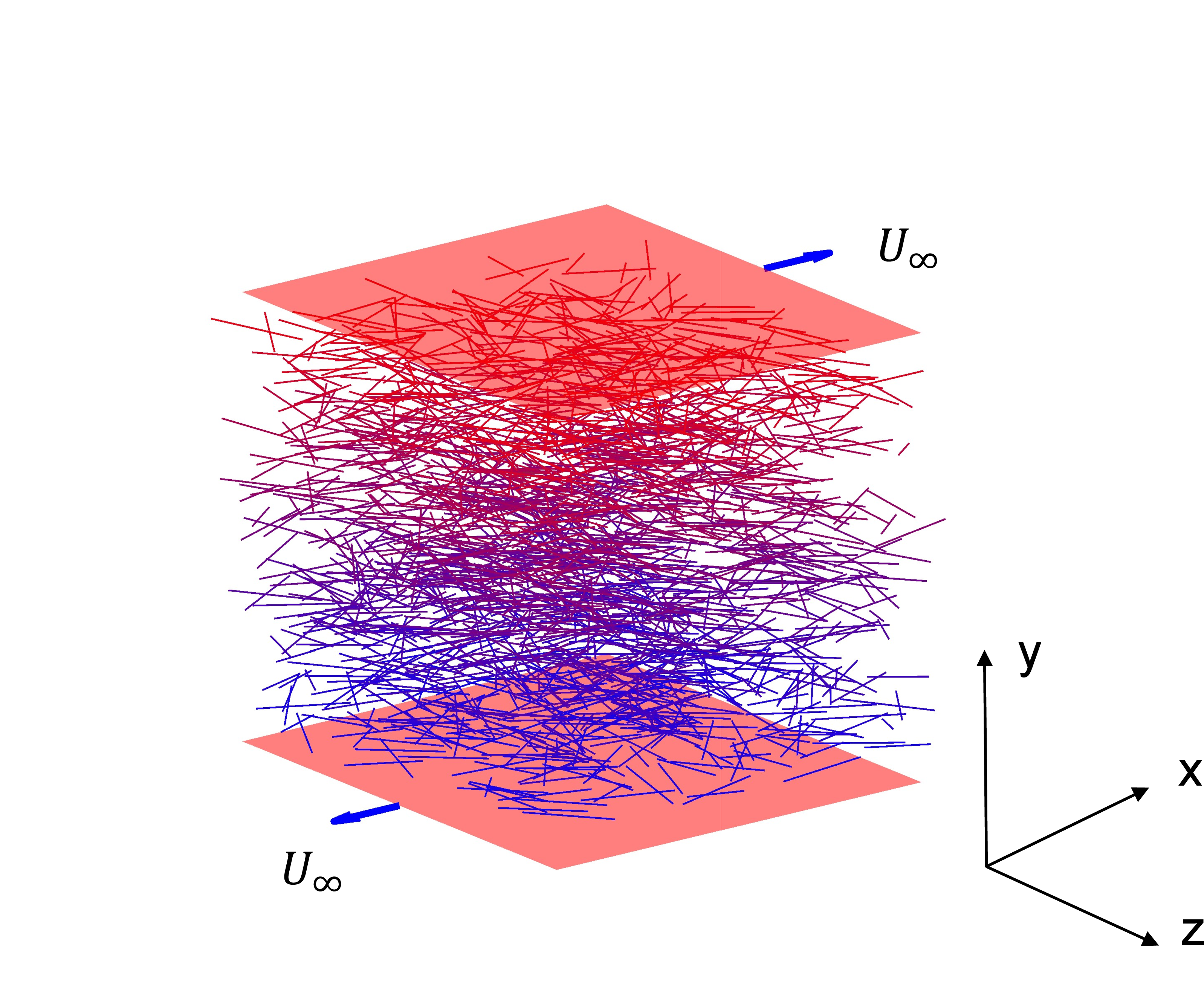}}
  \caption{Simulation setup of the shear flow of a fiber suspension. The top and bottom walls move with velocities $U_{\infty} = \dot{\gamma}L$ in the directions shown by the arrows.}
\label{fig:geometry}
\end{figure}

We investigate suspensions of flexible fibers in a Couette flow, by varying the volume fraction ($\phi$), aspect ratio ($AR = L/d$), flexibility (1/$\tilde{B}$), and roughness ($\epsilon_r$). The fibers are suspended in a channel with upper and lower walls moving in the x-direction with opposite velocities of magnitude $U_{\infty}=\dot{\gamma}L$. No-slip and no-penetration boundary conditions are imposed on the wall and periodicity is assumed in the stream-wise ($x$), and span-wise ($z$) directions. The fibers are initially distributed randomly. For this study, we consider a domain of size  $5L\times8L\times5L$ and $80\times128\times80$ grid points in the stream-wise ($x$), wall normal ($y$) and span-wise direction ($z$), respectively. The coordinate system and the simulation setup is illustrated in figure \ref{fig:geometry}. Simulations with bigger domain, higher grid and time resolution, e.g., 1.5, 2, 2.5 and 3 times the current domain, grid and time resolution show a negligible (less than 2 \%) change in the averaged steady state suspension viscosity. Finally, 17 Lagrangian points over the fiber length are enough to resolve the case with the highest fiber flexibility. The required time-step to capture the fiber dynamics is $\Delta t = 10^{-4}$. \textcolor{black}{The suspension is simulated until a statistically steady viscosity is observed and the mean values presented are obtained discarding the initial transients.}

\subsection{Stress and bulk rheology calculations}\label{sec:stress}

The bulk stress in the suspension quantifies the rheological properties of the suspension. \textcolor{black}{The total stress, ${\Sigma}_{ij}$, can be written as}

\begin{equation}
     {\Sigma}_{ij} = {\Sigma}{_{ij}^0}+{\Sigma}{_{ij}^f},
\end{equation}
\begin{equation}
     \begin{array}{l}
{\Sigma}{_{ij}^0}  = \frac{{{\mathop{\rm \textit{Re}}\nolimits} }}{V}\int\limits_{V - \sum {{V_f}} } {\left( { - p{\delta _{ij}} + \frac{2}{{{\mathop{\rm \textit{Re}}\nolimits} }}{e_{ij}}} \right)} dV,\\
\Sigma _{ij}^f = {\textit{Re} \over V}\sum \int\limits_{{V_f}} {{\sigma _{ij}}} dV - {{{\mathop{\rm \textit{Re}}\nolimits} } \over V}\int\limits_V {{{u'}_i}} {{u'}_j}dV.
\end{array}
\end{equation}
\textcolor{black}{Here ${\Sigma}_{ij}^0$ is the viscous fluid stress and results in a dimensionless contribution of 1 (or $\eta\dot{\gamma}$ in the dimensional form) in a simple shear flow after subtracting the isotropic fluid pressure. ${\Sigma}{_{ij}^f}$ is the stress generated by the presence of fibers and inter-fiber interactions. Moreover, $V$ is the total volume, $V_f$ the volume occupied by each fiber, ${e_{ij}} = \frac{{\partial {u_i}}}{{\partial {x_j}}} + \frac{{\partial {u_j}}}{{\partial {x_i}}}$ represents the strain rate tensor, $\boldsymbol{u}'$ the velocity fluctuations, and $\sigma_{ij}$ is the fiber stress.} The fiber stress $\sigma_{ij}$ can be decomposed into two parts:
\begin{equation}
\int\limits_{{V_f}} {{\sigma _{ij}}} dV = \int\limits_{{A_f}} {{\sigma _{ik}}} {x_j}{n_k}dA - \int\limits_{{V_f}} {\frac{{\partial {\sigma _{ik}}}}{{\partial {x_k}}}{x_j}} dV,
\end{equation}
where $A_f$ represents the surface area of each fiber and $\mathbf{n}$ is the unit surface normal vector on the fiber pointing outwards. The first term is called the stresslet, and the  second  term  indicates  the  acceleration  stress \citep{guazzelli2011physical}. The second term is identically zero for neutrally buoyant fibers when the relative acceleration of the fiber and fluid is zero. \textcolor{black}{Because $\sigma _{ik}{n_k}$ is simply the force per unit area acting on the fibers \citep{batchelor1971stress}, it can be written for slender bodies as:}

\begin{equation}
\int\limits_{{A_f}} {{\sigma _{ik}}} {x_j}{n_k}dA =  - r_p^2\int\limits_L { {F}_i} {x_j}ds,
\end{equation}
where ${F}_i$ is the fluid solid interaction force as defined in equation~\ref{eq:FSI_force}. The term $r_p^2$  arises from choosing the linear density instead of the volume density in the characteristic force scale. Finally, the total fiber stress is defined as:
\begin{equation}
    \Sigma {_{ij}^f}  = - \frac{{Rer_p^2}}{V}\sum\limits_1^n\int\limits_L {{F_i}{x_j}ds}  - \frac{{{\rm{\textit{Re}}}}}{V}\int\limits_V {{{u'}_i}} {{u'}_j}dV.
\end{equation}
From the results of our simulations, we observe that, the last term related to the velocity fluctuations is very small compared to the stresslet and can be neglected for the range of Reynolds number considered here. The calculated bulk stress tensor can now be used to quantify the rheological properties of the suspensions. The relative viscosity $\eta_r$ is defined as: 
\begin{equation}
     \eta_r = \frac{\eta_{eff}}{\eta},
\end{equation}
where $\eta_{eff}$ is the effective suspension viscosity. The relative viscosity in terms of the bulk shear stress is: 
\begin{equation}
     \eta_r  = 1 + {\Sigma}{_{xy}^f},
\end{equation}
with ${\Sigma}{_{xy}^f}$ being the ensemble averaged shear stress arising from the presence of the fibers, fluid-fiber, and inter-fiber interactions. The first ($N_1$) and second ($N_2$) normal stress differences are defined as:
\begin{subequations}
\begin{equation}
     N_1  = {\Sigma}{_{xx}}-{\Sigma}_{yy},
   \end{equation}
   \begin{equation}
     {N_2  = {\Sigma}{_{yy}}-{\Sigma}_{zz}}
\end{equation}
\end{subequations}

There are two main contributions to the bulk stress: 1) the hydrodynamic contribution $\Sigma_{ij}^h$, and 2) the contact contribution $\Sigma_{ij}^c$.
These will be useful in the following sections for analysing the simulation results. The contact contribution can be calculated from the ensemble average of the contact stresslet given by:

\begin{equation}
   \Sigma_{ij}^c =  -\frac{Rer_p^2}{V}\sum\limits_1^n \int_{L}{{F}^c_i} {x_j}ds.
\end{equation}
\textcolor{black}{Thus, the hydrodynamic contribution can be simply obtained as $\Sigma_{ij}^h=\Sigma_{ij}^0+({\Sigma}{_{ij}^f}-\Sigma_{ij}^c$).} This splitting of the total stress allows us to track the variations in the contributions from different mechanisms to the observed rheological behavior of the suspension with varying parameters, e.g., $\eta_r^h=\Sigma_{xy}^h$, and $\eta_r^c=\Sigma_{xy}^c$ are the hydrodynamic and the contact contributions to the relative viscosity $\eta_r$, respectively.
\subsection{Simulation parameters}
The aim of this study is to investigate the effects of dimensionless shear rate ($\dot{\Gamma}$),  fiber volume fraction ($\phi$), flexibility (1/$\tilde{B}$), aspect ratio ($AR$), and roughness ($\epsilon_r$) of the fibers on the suspension rheology in the presence of friction. First, we define the dimensionless numbers reported in this study. 

The relative importance of contact to the hydrodynamic forces can be estimated using dimensionless shear rate, $\dot\Gamma$ defined as 
\begin{equation}
    \dot {\Gamma}  = \frac{{6\pi \eta {d^2}\dot{\gamma}  }}{{{F_0}}},
\label{gamma}
\end{equation}
where $\dot\gamma$ is the imposed shear rate, and $\eta$ is the viscosity of the suspending fluid. To keep the Reynolds number $Re = \rho\dot{\gamma}L^2/\eta$ constant, we vary $\dot{\Gamma}$ by varying the characteristic contact force scale $F_0$. It is well known that the rate-dependent rheology of suspensions is determined by the competition between various stress (force) scales \citep{guazzelli2018rheology, more2020unifying}. Hence, by changing the value of $F_0$, we vary the relative contribution from the contact and hydrodynamic forces in the suspension. Typically, $F_0$ depends on the fiber material properties like Young's modulus, elastic modulus, Poisson's ratio, etc. \citep{brizmer2007elastic}. In some cases, this results in a very high value of $F_0$ restricting the time-step size to impractically small values ($< 10^{-7}$) making the simulations computationally expensive. \textcolor{black}{Hence, to resolve these numerical issues and gain some fundamental understanding}, we vary $F_0$ in orders of magnitudes of the hydrodynamic stress scale, i.e., $6{\pi}{\eta}{d^2}\dot{\gamma}$, as routinely done for spherical suspensions \citep{gallier2014rheology, mari2014shear, more2020roughness}. This automatically allows us to vary $\dot{\Gamma}$, without changing the Reynolds number. The dimensionless roughness height $\epsilon_r$ is defined as  

\begin{equation}
    \epsilon_r = \frac{h_r}{d}
\end{equation}
where $h_r$ is the surface roughness height. The fiber bending rigidity $\tilde{B}$, which quantifies the fiber flexibility is made dimensionless by the viscous stress scale as: 
\begin{equation}
  \tilde{B} = \frac{EI}{\eta\dot{\gamma}L^4}
\end{equation}

\textcolor{black}{In the literature, the fiber rigidity is quantified using the dimensionless bending rigidity ($\tilde{B}$) \citep{switzer2003rheology, keshtkar2009rheological,bounoua2016shear}, and/or bending ratio ($BR$) \cite{wu2010numerical,snook2014normal}. From their respective definitions, one can see that they can be used interchangeably. The Bending ratio,  $BR$ $\Bigl(BR=\frac{\tilde{B}(ln2r_e-1.5)*32}{\pi}\Bigl)$ is just a multiple of the dimensionless bending rigidity, $\tilde{B}$. So, in this study, effects of varying the dimensionless bending rigidity are equivalent to that observed by varying the bending ratio.}  
This dimensionless bending rigidity $\tilde{B}$ is related to the bending rigidity $B$ reported in equation (\ref{non-dimensional_fil}) by 
\begin{equation}
    \tilde{B} = \frac{\pi }{4}r_p^2{\mathop{\rm \textit{Re}}\nolimits} B.
\end{equation}
The volume fraction of the suspension is defined as
\begin{equation}
    \phi  = \frac{{n\pi r_p^2}}{{4V}}.
\end{equation}

In summary, the functional form of the stress can be written as: : 

\begin{eqnarray}
    \frac{\Sigma_{xy}}{\eta\dot{\gamma}} = \eta_r(\frac{{n\pi r_p^2}}{{4V}},\frac{L}{d},\frac{h_r}{d},\frac{EI}{\eta\dot{\gamma}L^4}, \frac{{6\pi \eta {d^2}\dot{\gamma}  }}{{{F_0}}})\nonumber\\
    =
    \eta_r (\phi, AR, \epsilon_r,\tilde{B}, \dot{\Gamma})
\end{eqnarray}
We simulate a shear flow by varying the fiber bending rigidity in the range $0.005\leq \tilde{B} \leq0.2$, the dimensionless roughness height in the range $0.005\leq \epsilon_r\leq0.10$, the dimensionless shear rate in the range $1\leq \dot\Gamma\leq1000$, and the aspect ratio in the range $10 \leq AR \leq 16$. Note that the lower (higher) $\tilde{B}$ value corresponds to flexible (rigid) fibers. The range of parameters explored in the present work are summarized in table \ref{tab:my-table}.
\begin{table*}
\caption{Range of parameters explored in this study}
 \begin{tabular}{lcccc}
      $\epsilon_r$   &   $\dot\Gamma$ & $\phi$  & $AR$ & $\tilde{B}$\\[10pt]
      \hline\\
       $0.005-0.10$~ & ~~$1-1000$~ &~~ $0.05-0.46$~ &~~$10-16$~ &~~$0.005-0.2$\\
  \end{tabular}
  \label{tab:my-table}
\end{table*}

\section{Results and discussion}

This section presents the results of the numerical simulations. Sections \ref{shear rate_behavior}-\ref{sec:flex} focus on the shear rate dependent behavior and explain the effects of varying the fiber volume fraction, aspect ratio, flexibility, and roughness. Section \ref{sec:normal} presents the effects of the governing parameters on the normal stress differences $N_1$ and $N_2$.  

\begin{figure}
\centering

\includegraphics[width=0.5\linewidth]{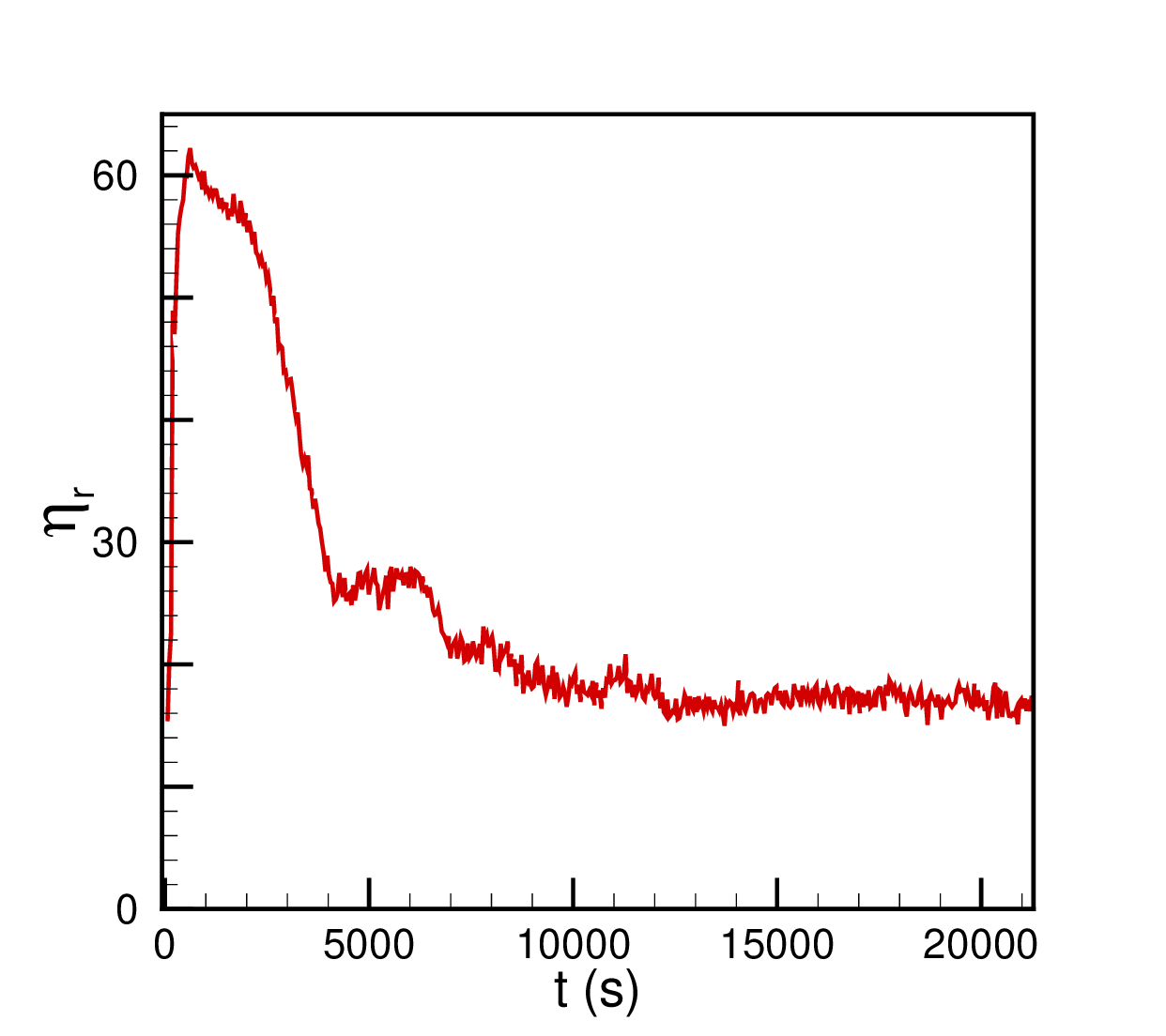}

\caption{Transient viscosity from the initial state to the steady state flow. The roughness height is fixed to $\epsilon_r$ = 0.05, bending rigidity to $\tilde{B} = 0.07$, aspect ratio to $AR$ = 16, and volume fraction to $\phi = 15\%$.  }\label{fig:transient}
\end{figure}

We make all the quantities dimensionless using the scaling mentioned in section \ref{model}. The suspension is simulated until a statistically steady viscosity is observed and the mean values presented are obtained discarding the initial transients as shown in figure~\ref{fig:transient}
\textcolor{black}{The structural solver has been validated by comparing four distinct cases in a previous study \cite{alizad2017numerical} and hence, not repeated here for brevity.}

\subsection{Shear-thinning and increase of $\eta_r$ with the volume fraction}\label{shear rate_behavior}

\begin{figure*}
\centering
\begin{subfigure}{.5\textwidth}
  \centering
\begin{overpic}[width=1.0\linewidth]{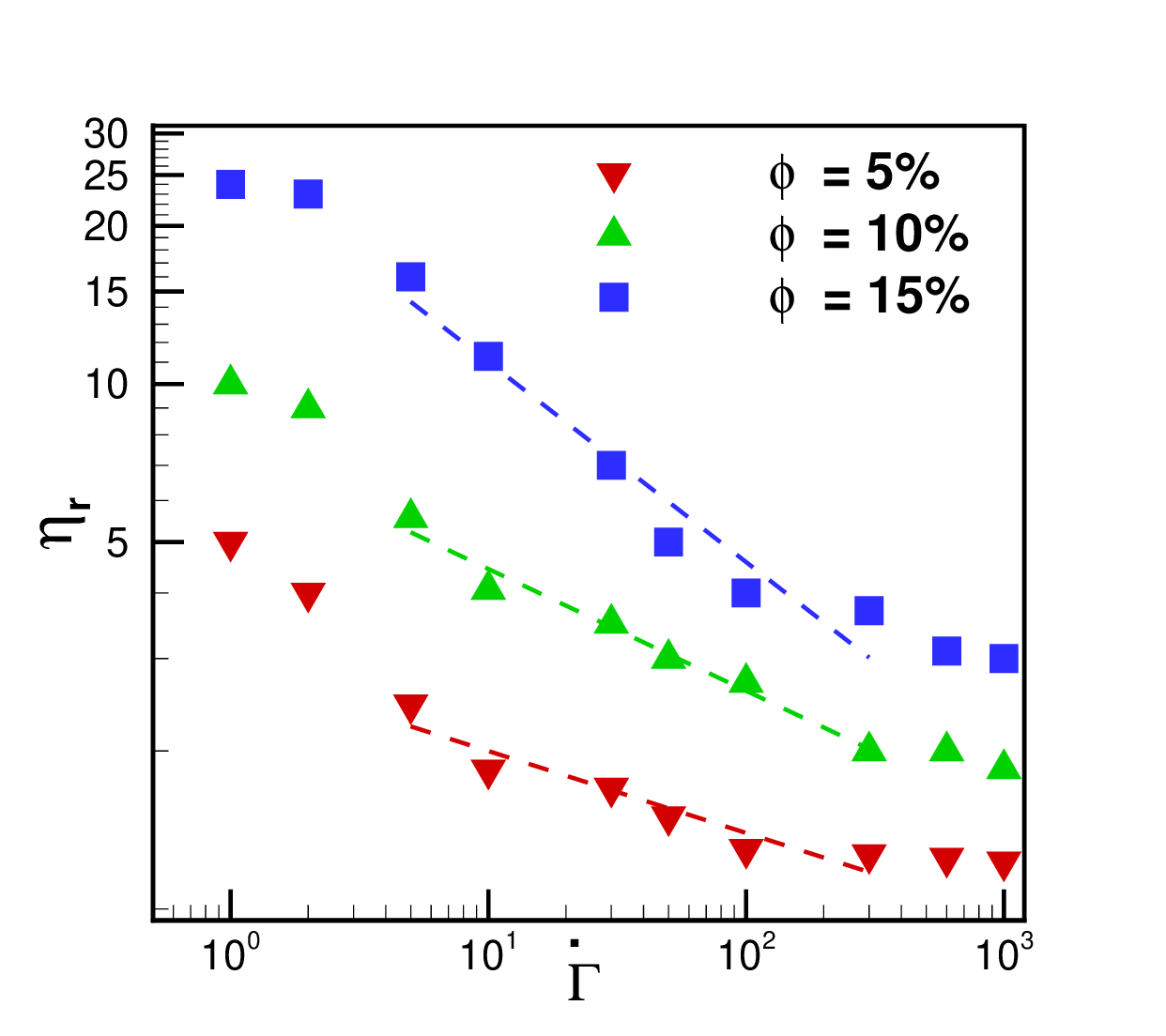}
 \put (42,52) {$\textcolor{black}{{\boldsymbol{\eta_r  = 26.47\dot {\Gamma}^{-0.381}}}}$}
 \put (46,25) {$\textcolor{green}{{\boldsymbol{\eta_r  = 7.58\dot {\Gamma}^{-0.232}}}}$}
 \put (14,15) {$\textcolor{red}{{\boldsymbol{\eta_r  = 2.86\dot {\Gamma}^{-0.156}}}}$}
\end{overpic}
   \caption{}
  \label{fig:Shear_volume}
\end{subfigure}%
\begin{subfigure}{.5\textwidth}
  \centering
  \includegraphics[width=1.0\linewidth]{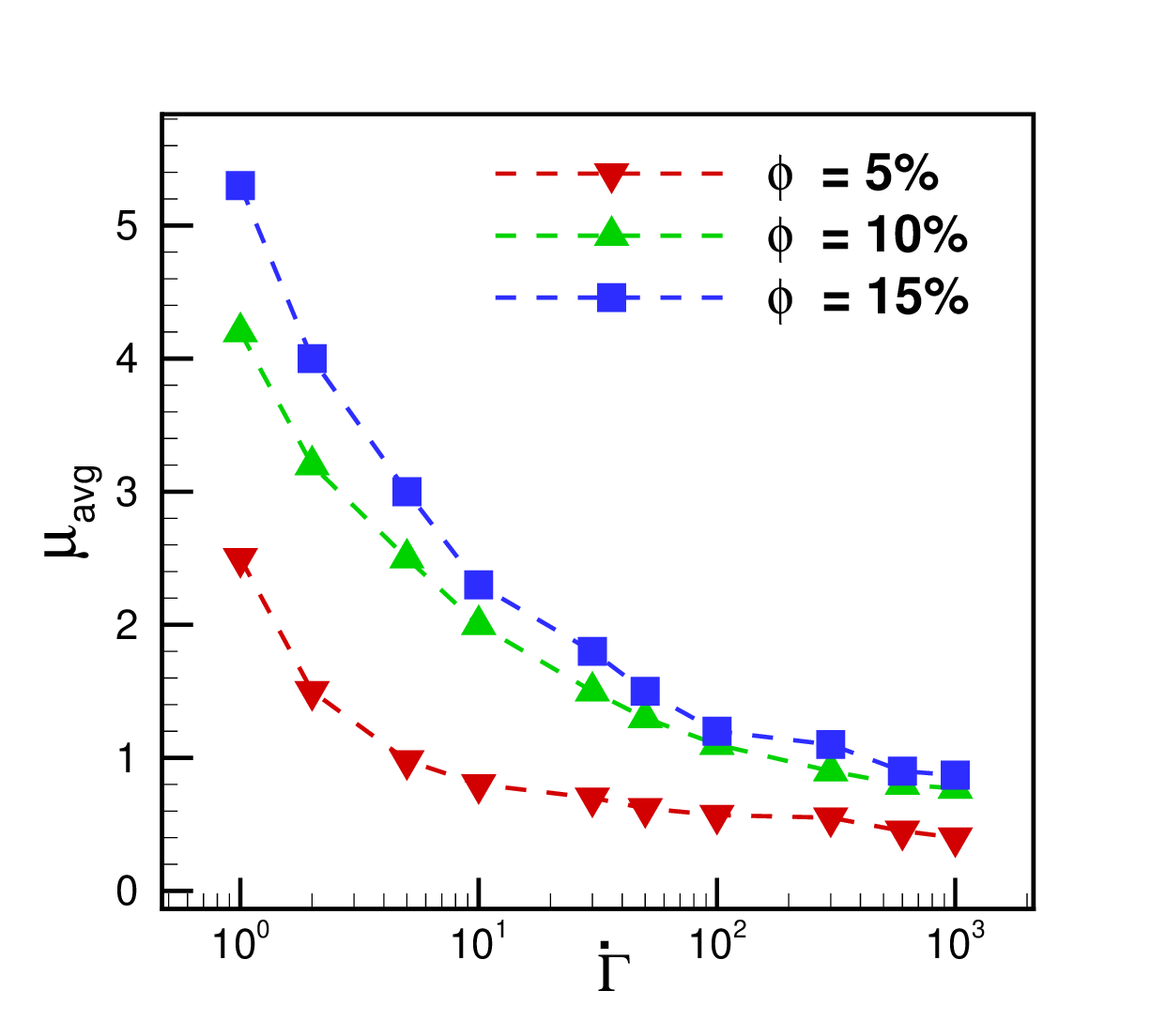}
  \caption{}
  \label{fig:mu_shear_vol}
\end{subfigure}
\caption{(a) Relative viscosity $\eta_r$ of fiber suspensions versus the dimensionless shear rate $\dot{\Gamma}$ at different volume fractions. The dashed line represents the power law fit as described in equation \ref{power_law}. The power law indices for different volume fractions are - $\phi = 5\%$: $k = 2.86$, $ m = 0.156 $; $\phi = 10\%$: $k = 7.58$, $m = 0.232$; and $\phi = 15\%$: $k = 26.47$, $m = 0.381 $. (b) Variation in the average coefficient of friction $\mu_{avg}$ for the suspensions with dimensionless shear rate for different volume fractions.  The roughness height is fixed to $\epsilon_r$ = 0.05, bending rigidity to $\tilde{B} = 0.07$ and aspect ratio to $AR$ = 16 for all cases. The  model implemented in this study can reproduce the experimentally observed shear thinning behavior in suspensions of fibers.}
\label{fig:shear_volume_rough}
\end{figure*}

\begin{figure*}
\centering
\begin{subfigure}{.5\textwidth}
  \centering
  \includegraphics[width=1.0\linewidth]{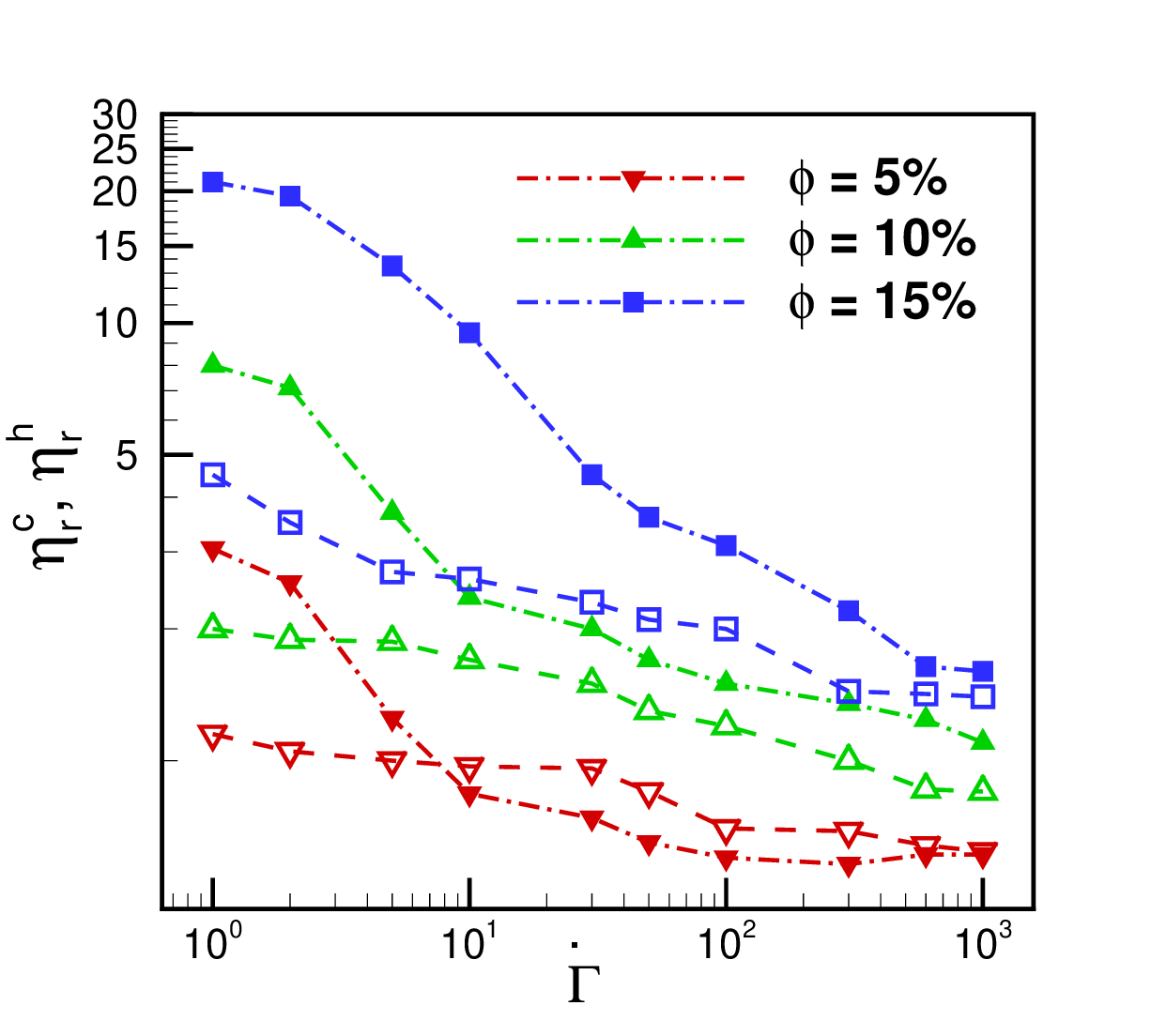}
  \caption{}
  \label{fig:shear_rate_stress_budget}
\end{subfigure}
\begin{subfigure}{0.5\textwidth}
  \centering
  \includegraphics[width=1.0\linewidth]{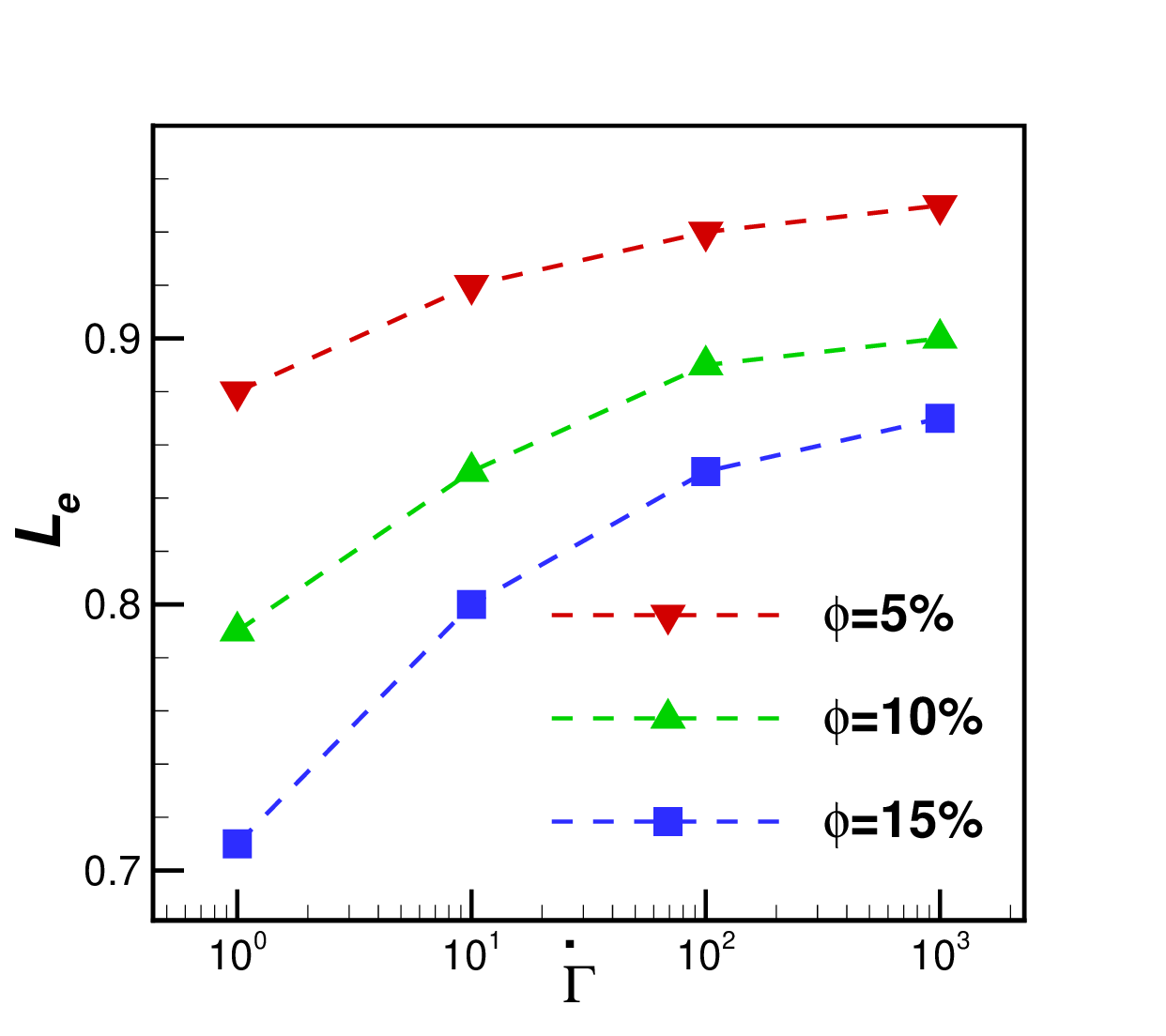}
  \caption{}
  \label{fig:end_to_end_shear}
\end{subfigure}
\caption{(a) Hydrodynamic $\eta_r^h$ and contact $\eta_r^c$ contributions to the relative
viscosity for three different volume fractions as in figure~\ref{fig:Shear_volume}. Open symbols: hydrodynamic contribution. Closed symbols: contact contribution. \textcolor{black}{The sum of the hydrodynamic and contact contribution is the total relative viscosity as reported in figure \ref{fig:Shear_volume}.} (b) Ensemble averaged end-to-end distance $L_e$ of the suspended fibers as a function of dimensionless shear rate for different volume fractions.  The roughness height is fixed to $\epsilon_r$ = 0.05, bending rigidity to $\tilde{B} = 0.07$ and aspect ratio to $AR$ = 16 for all cases. }
\label{fig:sharratebudget}
\end{figure*}

We start with analyzing the computed relative viscosity, $\eta_r$ as a function of the dimensionless shear rate, $\dot{\Gamma}$, for different volume fractions, as shown in figure~\ref{fig:Shear_volume}. We observe that the viscosity reduces with increasing the shear rate, 
indicating a shear thinning behavior. Eventually, it plateaus at high shear rates.
To quantify the degree of shear thinning, we use the following power law curve fit to the simulation data.
\begin{equation}
    \eta_r(\dot {\Gamma})  = k\dot {\Gamma}^{-m}.
\label{power_law}
\end{equation}
Here $k$ is the flow consistency index and $m>1$ is the shear thinning index, which quantifies the strength of the shear thinning effect. As shown in figure \ref{fig:Shear_volume}, with increasing the volume fraction of the suspension, the magnitudes of $k$ and $n$ increase, indicating a rise in the shear thinning effect with increasing the volume fraction.

\begin{figure}
\centering

\includegraphics[width=0.5\linewidth]{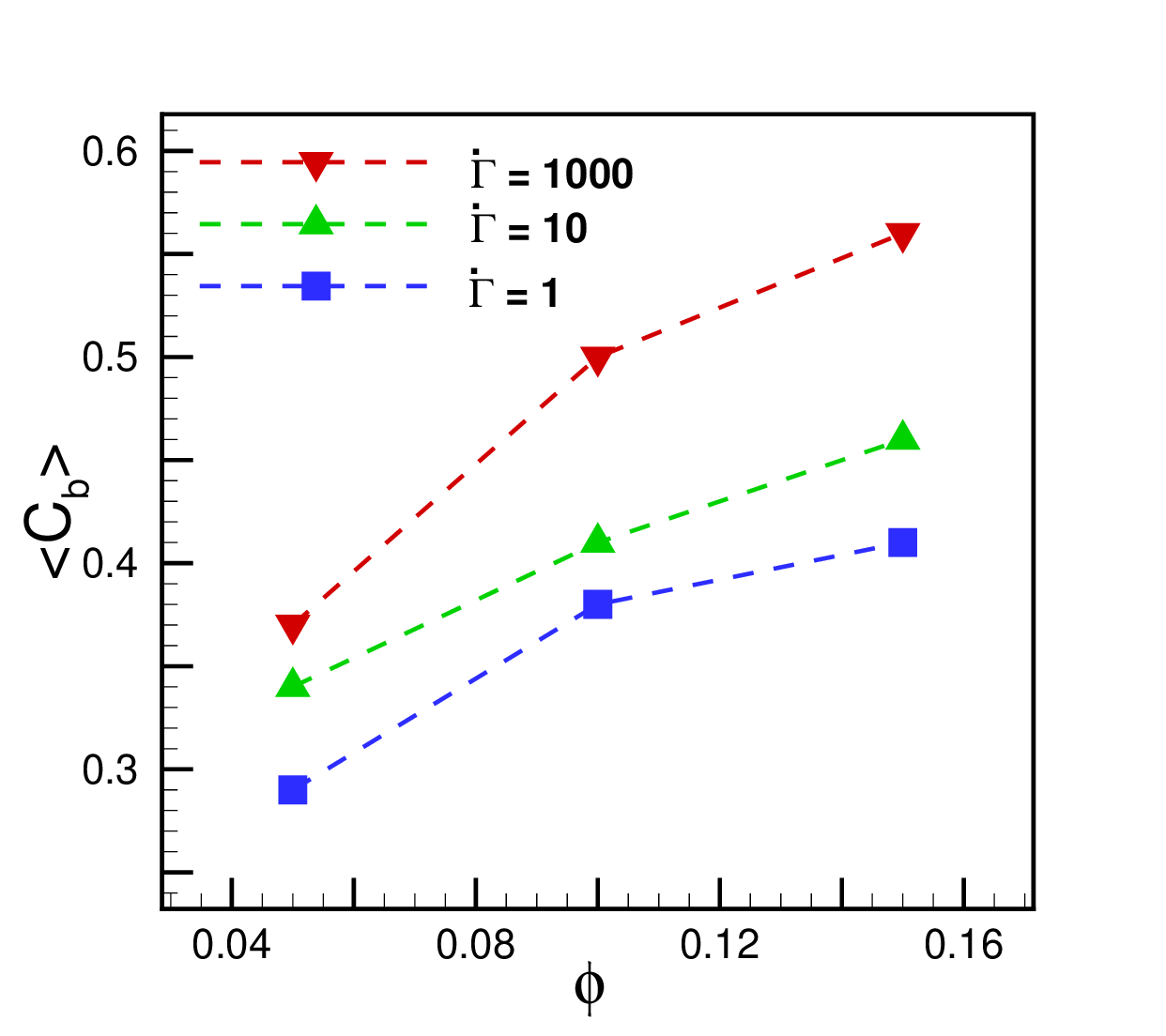}

\caption{The mean orbit constant $<C_b>$ for different volume fraction. As the volume fraction increase, fibers align more strongly in the flow gradient plane.}\label{fig:C_b_volume}
\end{figure}

\textcolor{black}{Increase in the shear stress directly increases the magnitude of the normal force between the fibers, which in turn results in the reduction of the friction coefficient according to equation~\ref{eq:frictionlaw}. Hence, with all other parameters maintained constant, we expect a drop in the coefficient of friction with increasing shear rate (which is directly proportional to the shear stress), leading to a decrease in the relative viscosity, similar to the findings of More \textit{et al.} \cite{more2020roughness}.} This becomes clear when we look at  the average coefficient of friction ($\mu_{avg}$) as depicted in figure~\ref{fig:mu_shear_vol}. $\mu_{avg}$ is the time average of the coefficient of friction, $\mu$, between all the contacting pairs in the suspension when the simulations reach statistically steady state. This reduction in $\mu_{avg}$ with $\dot{\Gamma}$ results in a decrease in the dominant contact contribution $\eta_r^c$ to the bulk suspension viscosity as depicted in figure~\ref{fig:shear_rate_stress_budget}. \textcolor{black}{ On the other hand, a constant friction coefficient results in a constant shear rate independent viscosity as showed in our previous publication \cite{khan2022constitutive}.} Moreover, the hydrodynamic contribution $\eta_r^h$ to the relative viscosity also decreases with $\dot{\Gamma}$ adding to the shear thinning. We can explain the reduction of the hydrodynamic contribution with shear rate by calculating the fiber deformation as reported in figure \ref{fig:end_to_end_shear}.
As the shear rate increases, fibers' deformation decreases and they align more with the flow direction, eventually decreasing the hydrodynamic contributions. 
\textcolor{black}{Moreover, the orbit constant, $C_b$, when averaged over the fibers in the suspension, is the measure of the orientation distribution. The ratio $C/(C+1)$ is used to define $C_b$ where,}
\begin{equation}
    C = \frac{1}{A}\tan \theta_z {({{\rm A}^2}{\cos ^2}\theta_y  + {\sin ^2}\theta_y )^{1/2}}
\end{equation}
\textcolor{black}{where $\theta_z$ is the angle of the fiber with respect to the span-wise direction (z-axis), and angle $\theta_y$ is the  angle between the gradient direction (y-axis) and the projection of the fiber on the flow-gradient (x-y) plane. The average orbit  constant$<C_b>$ goes  to 1  if  all  fibres  are  aligned  in  the  flow–gradient  plane  and  equals  zero  if the  fibres  are  aligned  with  the  span-wise direction (z-axis). Figure~\ref{fig:C_b_volume} shows that as the volume fraction increases, fibers align more strongly to the flow gradient plane, similar to the findings of previous simulations \cite{wu2010numerical,lindstrom2008simulation, snook2014normal}. Moreover, as the shear rate increases, fibers also tend to align with the flow-gradient direction}.

\begin{figure*}
\centering
\begin{subfigure}{.5\textwidth}
  \centering
  \includegraphics[width=1.0\linewidth]{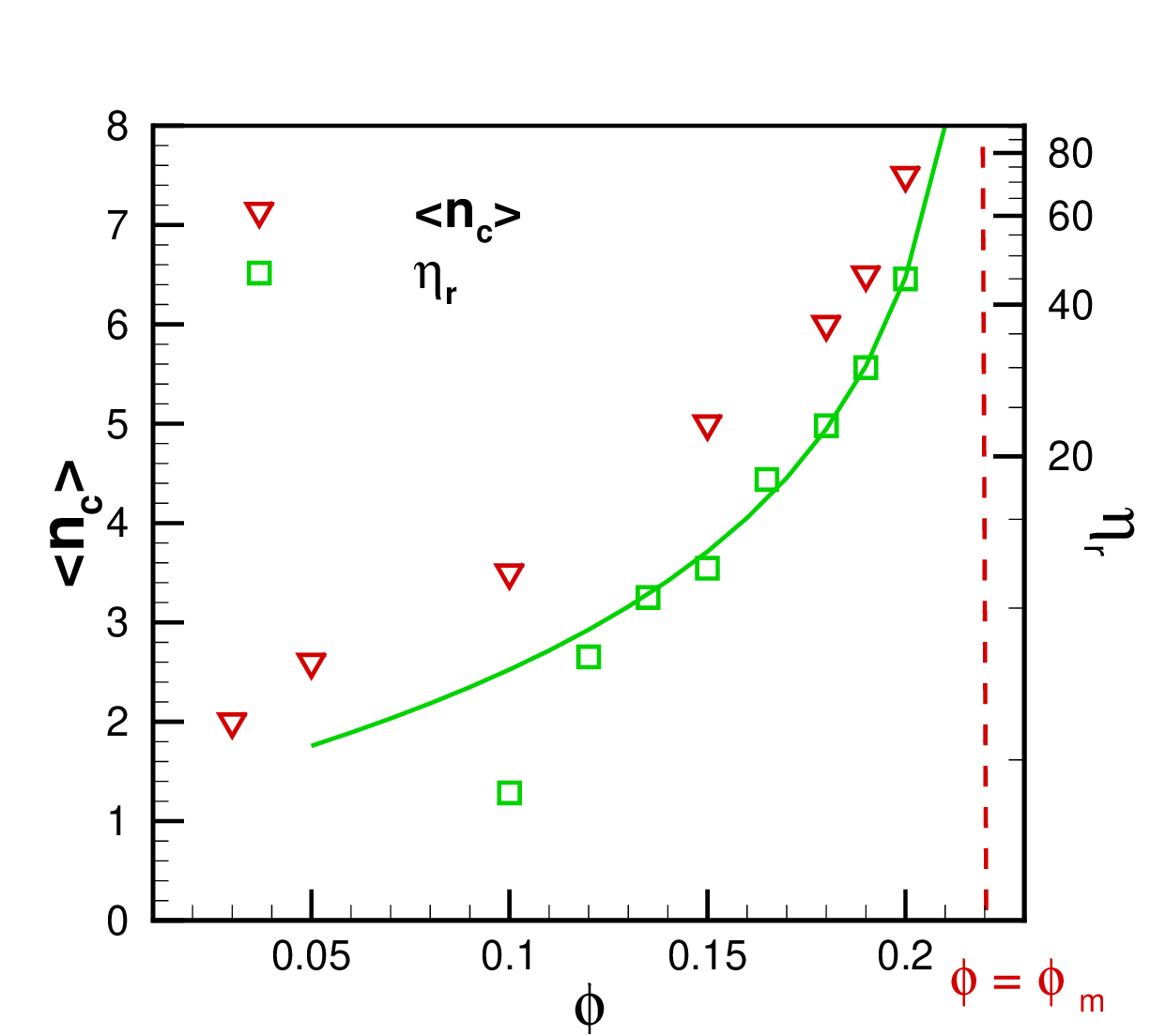}
  \caption{}
  \label{fig:contact_volume}
\end{subfigure}
\begin{subfigure}{0.5\textwidth}
  \centering
  \includegraphics[width=1.0\linewidth]{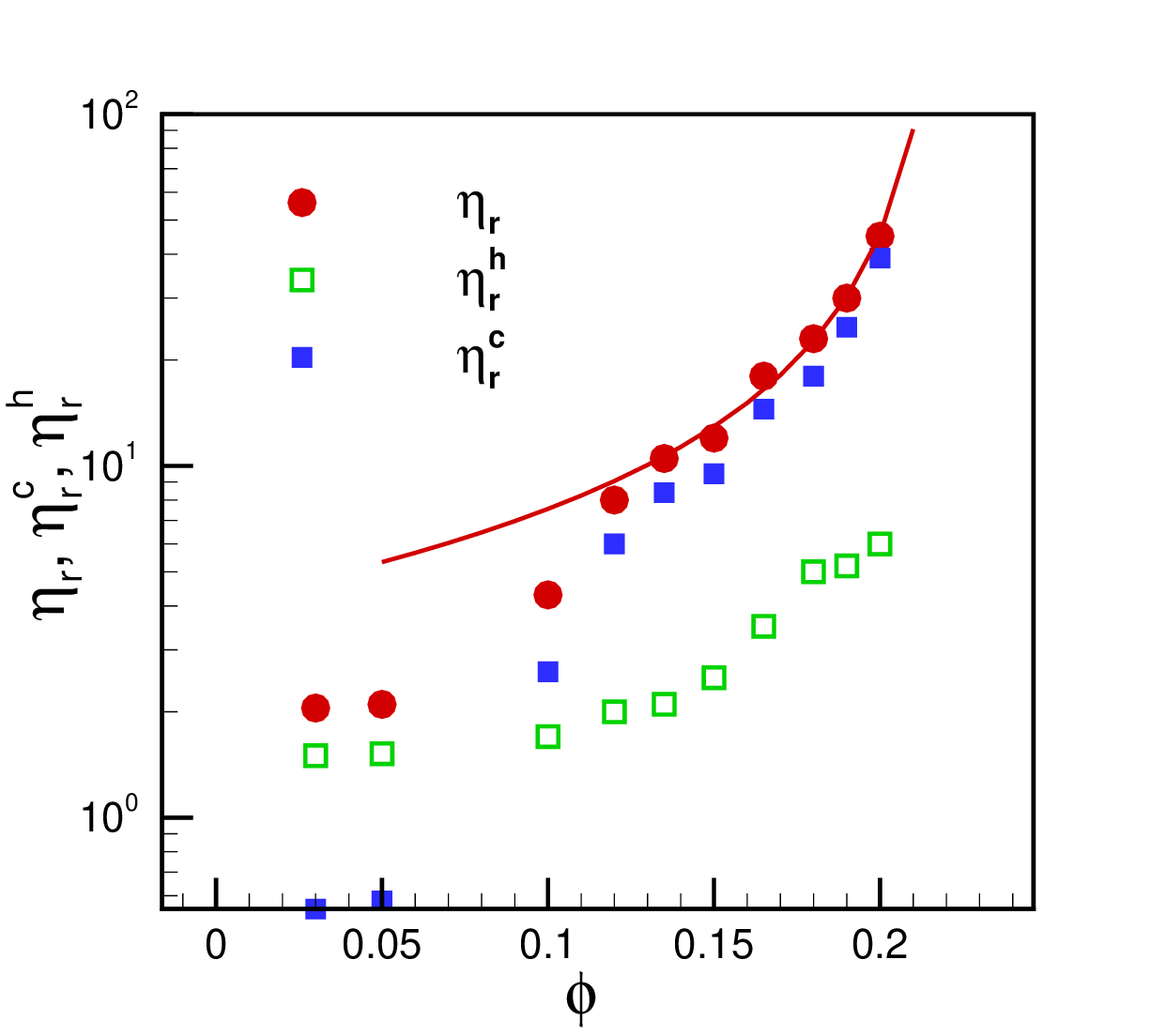}
  \caption{}
  \label{fig:stress_budget_volume}
\end{subfigure}
\caption{(a) The average number of contacting fibers and viscosity variation vs. volume fraction. (b) Hydrodynamic and contact contributions to the relative viscosity as a function of volume fraction. The average number of contacting fibers increases with the volume fraction which in turn increases the relative contribution from the contact interactions to the total viscosity. The green solid line in (a) represents the fitting with a modified Maron-Pierce law  \citep{maron1956application} described as 
$n_r=\alpha (1- \phi/\phi_m)^{-\beta}$, with parameters $\alpha = 4.42$, $\phi_m = 0.22 $  and $\beta = 1$. The dashed red line shows the jamming volume ($\phi_m = 0.22$) for the current case where roughness height was fixed to $\epsilon_r$ = 0.05, shear rate to $\dot {\Gamma} = 10$, bending rigidity to $\tilde{B} = 0.07$, and aspect ratio to $AR = 16$.}
\label{fig:contact_contri}
\end{figure*}

\textcolor{black}{It is well known that the viscosity of suspensions increases with increasing the particle phase volume fraction and it diverges at a certain maximum volume fraction beyond which the suspension does not flow, the so-called the jamming fraction $\phi_m$}. We observe the same behavior for fiber suspensions in our simulations. With all other parameters held fixed, the suspension viscosity increases with increasing the volume fraction as depicted in figure~\ref{fig:contact_volume}. The reason behind this increase is the increase in the fiber contribution to the total stress with increase in their numbers per unit volume. In our simulations, we quantify this contribution using $\Sigma_{xy}^f$ as discussed in sec.~\ref{sec:stress}. The fiber contribution to the total stress increases with the volume fraction due to the increase in the average number of contacting fibers ($<n_c>$) as also depicted in figure~\ref{fig:contact_volume}. \textcolor{black}{$<n_c>$ is the time average of the average number of fibers that come into contact with another fiber once the simulations reach statistically steady state.} The consequence of the increase in $<n_c>$ with $\phi$ is the rise in the contribution of the hydrodynamic as well as contact interactions to the total viscosity as shown in figure~\ref{fig:stress_budget_volume}. While the hydrodynamic contribution increase weakly, the contact contribution becomes significant and dominant at volume fraction greater than 5\% because of the substantial rise in $<n_c>$. \textcolor{black}{This observation underlines the impact of contacts in the rheology of fiber suspensions in the semi-dilute and the concentrated regime.} 

The variation in $\eta_r$ with $\phi$ can be captured by the the well known Maron-Pierce law \citep{maron1956application}, which has been extensively used in the literature to model the behavior of smooth spherical particle suspensions.
However, to capture the behavior of rough suspensions, a modified Maron-Pierce law is needed \citep{more2020effect}. \textcolor{black}{The following modified Maron-Pierce law \citep{maron1956application, more2020effect} is used to fit the data in the figure \ref{fig:contact_volume}:}

\begin{equation}
    n_r=\alpha (1- \phi/\phi_m)^{-\beta},
    \label{fitting}
\end{equation}
with the coefficient $\alpha$, the exponent $\beta$, and the jamming fraction $\phi_m$ depending on the aspect ratio and roughness of the fibers. As depicted by the solid lines in figure  \ref{fig:contact_volume}, equation \ref{fitting} does a good job in modelling the viscosity as a function of volume fraction.  We find that the relative viscosity diverges near the jamming transition with a scaling of $(\phi_m - \phi)^{-1}$ in contrast with $(\phi_m - \phi)^{-2}$, the behavior observed for a suspension of spheres \citep{more2020constitutive, more2020roughness, more2020effect}. The scaling obtained here for flexible fibers is $(\phi_m - \phi)^{-1}$, whereas for rigid fibers the scaling was found as $(\phi_m - \phi)^{-0.90}$ \cite{tapia2017rheology}.  In the next subsection, we turn our attention to the effect of the fiber aspect ratio on suspension rheology.


\subsection{Increase in aspect ratio of the fibers reduces the suspension flowability}

\begin{figure*}
\centering
\begin{subfigure}{.5\textwidth}
  \centering
  \includegraphics[width=1.0\linewidth]{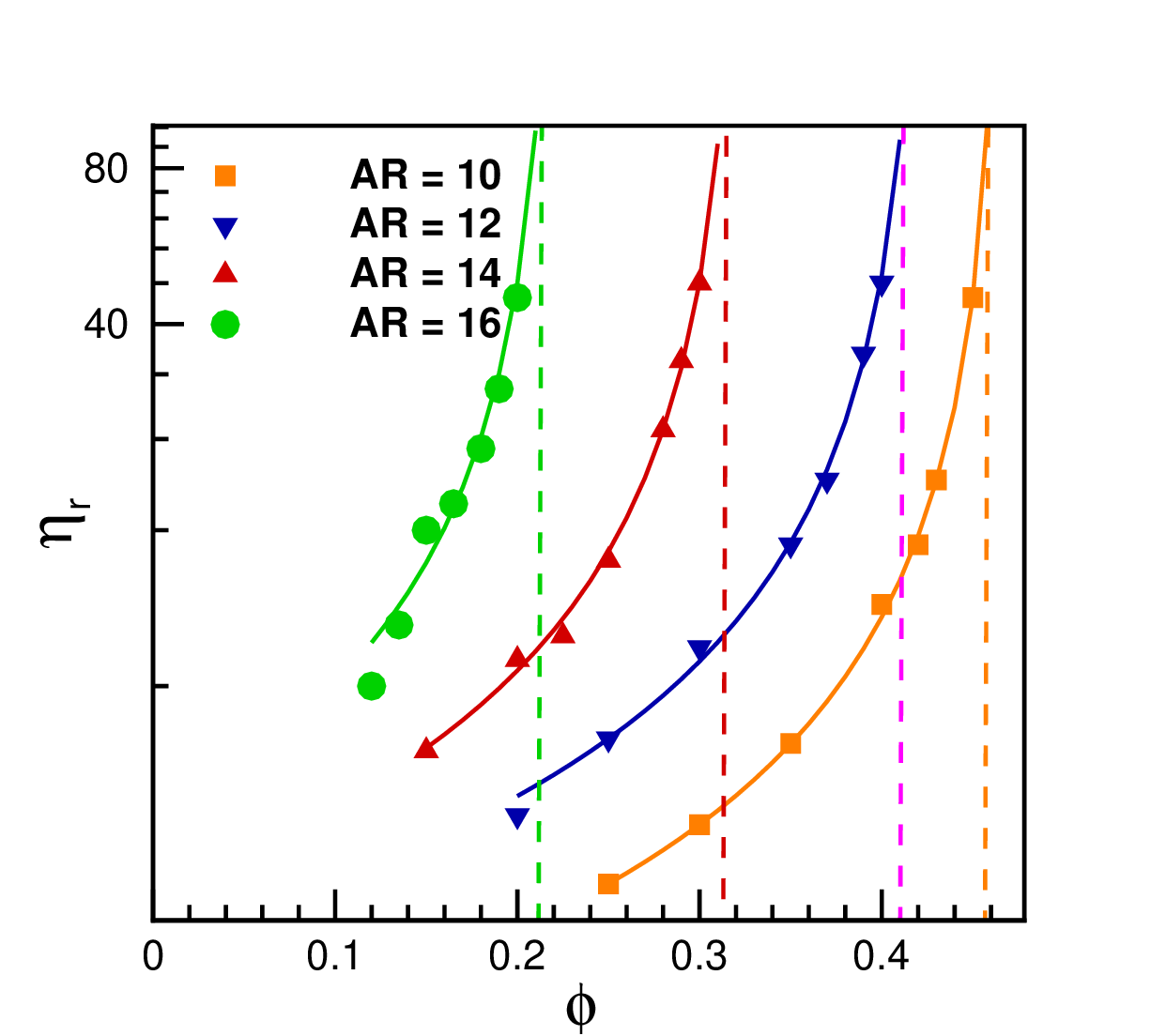}
  \caption{}
  \label{fig:ARjamming}
\end{subfigure}%
\begin{subfigure}{.5\textwidth}
  \centering
  \includegraphics[width=1.0\linewidth]{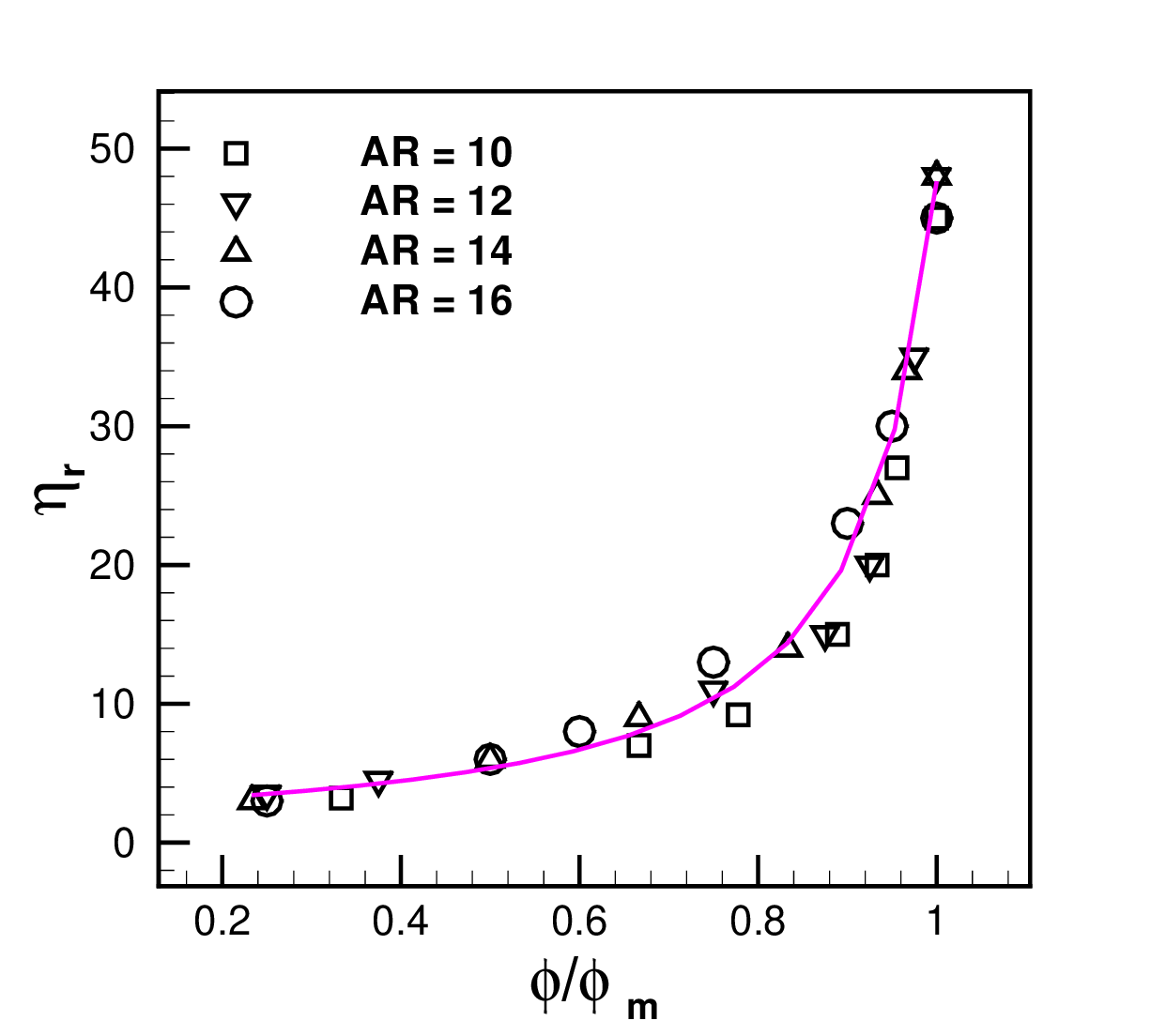}
  \caption{}
  \label{fig:AR_sameplot}
\end{subfigure}
\caption{(a) Relative viscosity $\eta_r$ vs. volume fraction $\phi$ for different aspect ratios $AR$. Solid lines are the fitting curve using equation \ref{fitting}. The vertical dashed
line shows the jamming volume fraction for different aspect ratios. (b) Re-scaled rheological data for different aspect ratios. All simulations were performed for a fixed  roughness $\epsilon_r = 0.05$ and bending rigidity, $\tilde{B} = 0.07$. The jamming fraction reduces with aspect ratio for a given shear rate. Re-scaling the volume fraction by $\phi_m$ leads to a collapse of the viscosity curves, indicating that the aspect ratio determines the maximum volume fraction once the fiber flexibility and roughness are fixed.}
\label{fig:AR_jamming_Full}
\end{figure*}

As shown in figure~\ref{fig:contact_volume}, the average number of contact for a fiber increase with increasing volume fraction. The same is true with increasing the fiber aspect ratio, $AR$ \citep{williams2003}. In a concentrated suspension, elongated fibers have a larger probability of inter-fiber contacts. As a result, there is an increase in the contact force which leads to a caging effect. Thus, the viscosity increases with the aspect ratio for a fixed volume fraction as shown in figure \ref{fig:ARjamming}. As a result, the viscosity diverges at a critical volume fraction $\phi_m$ that depends on the aspect ratio as shown in figure \ref{fig:ARjamming}. The solid lines demonstrate the best fit with equation~(\ref{fitting}).
The values for $\alpha$ and $\phi_m$ obtained after fitting are presented in table \ref{tab:jamming_AR}. From the data, we observe a reduction of $\phi_m$ from $\approx 0.47$ to $\approx 0.22$ with increasing the fiber $AR$ from $10$ to $16$ for suspensions with  $\epsilon_r=0.05$, and $\tilde{B}=0.07$. Figure \ref{fig:AR_sameplot} shows the data after rescaling the volume fraction by $\phi_m$, which leads to a collapse of the data indicating that it is the distance of $\phi$ from the jamming fraction $\phi_m$ that determines the suspension rheology. Hence, $\phi/\phi_m$ can be used as a design parameter for tuning the fiber suspension rheology, as it is easier to measure and control compared to other parameters like size distribution, roughness and friction in real-world suspensions. In the next subsection, we focus on the effect of roughness of the fibers on the suspension rheology. 

\begin{table}
    \caption{Fitting parameters for different aspect ratios for a fixed roughness height, $\epsilon_r =0.05$, bending rigidity, $\tilde{B} = 0.07$ and dimensionless shear rate, $\dot{\Gamma}=10.0$ . $\beta = 1$ for all cases.}
\def~{\hphantom{1}}
  \begin{tabular}{lcc}
      $AR$  & $\alpha$   &   $\phi_m$   \\
      \hline\\
       $10$   &~ $1.54$~ & ~$0.47$~ \\
       $12$   & ~$2.58$~ & ~$0.42$~ \\
       $14$  & ~$3.24~$ & ~$0.32$~ \\
      $ 16$   & ~$4.42$~ & ~$0.22$~ \\
       
  \end{tabular}
  
  \label{tab:jamming_AR}
\end{table}

\subsection{Roughness increases suspension viscosity}\label{sec:roughness}

\begin{figure*}
\centering
\begin{subfigure}{.5\textwidth}
  \centering
  \includegraphics[width=0.7\linewidth]{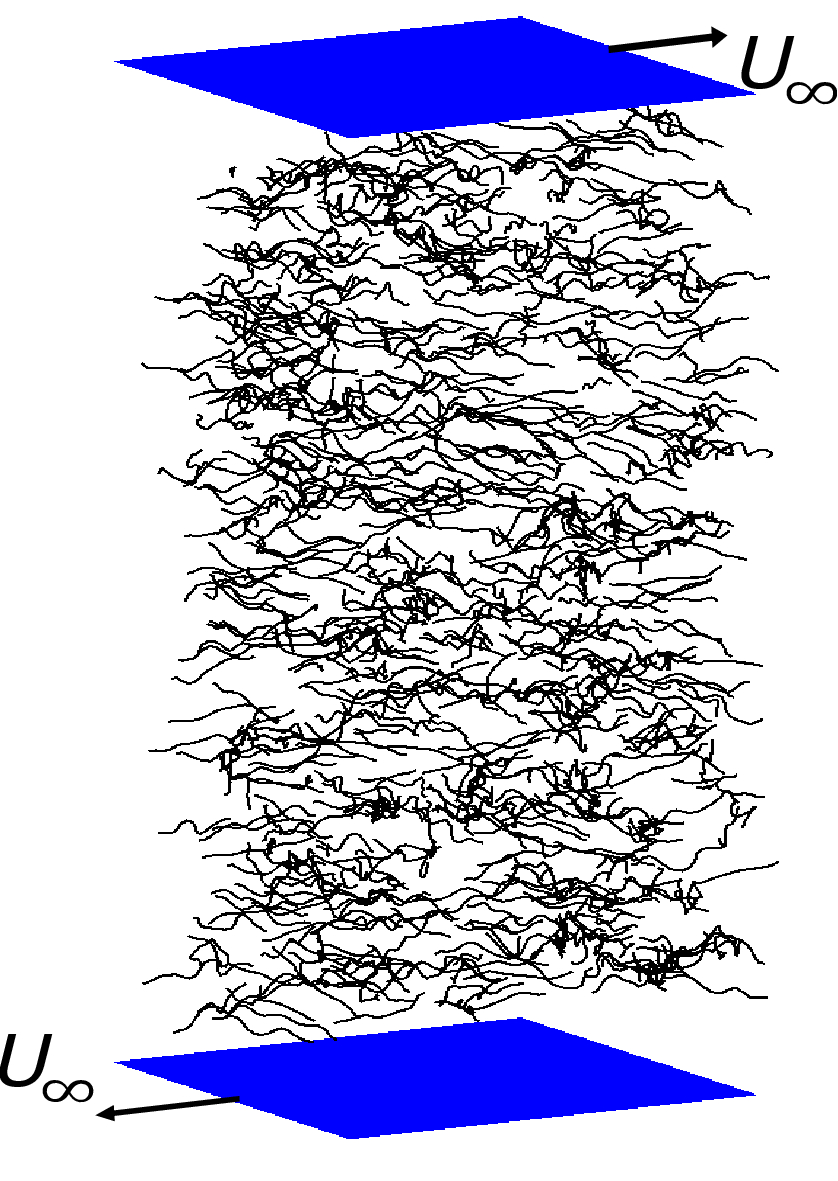}
  \caption{}
  \label{fig:visulaize_.005}
\end{subfigure}%
\begin{subfigure}{.5\textwidth}
  \centering
  \includegraphics[width=0.7\linewidth]{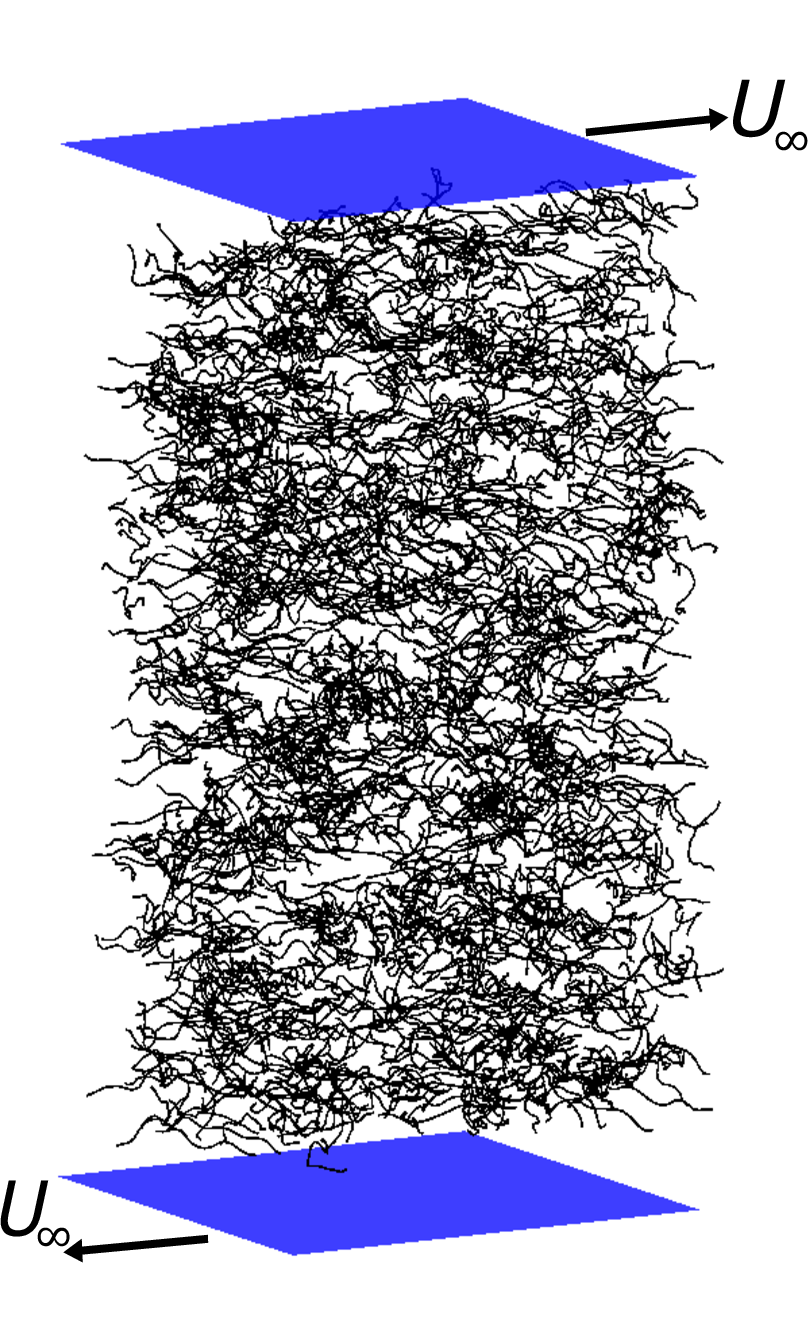}
  \caption{}
  \label{fig:visulaize_.20}
\end{subfigure}
\caption{Simulation snapshots at a thin slice in the middle of the simulation cell after shearing  to a shear rate of $\dot{\Gamma}=10$ with bending rigidity, $\tilde{B}=0.07$, aspect ratio, $AR = 16$, and roughness height, (a) $\epsilon_r = 0.005$, (b) $\epsilon_r = 0.05$. Rough fibers tend to agglomerate more compared to smooth fibers.}
\label{fig:visual_3D}
\end{figure*}

\begin{figure*}
\centering
\begin{subfigure}{.5\textwidth}
  \centering
  \includegraphics[width=1.0\linewidth]{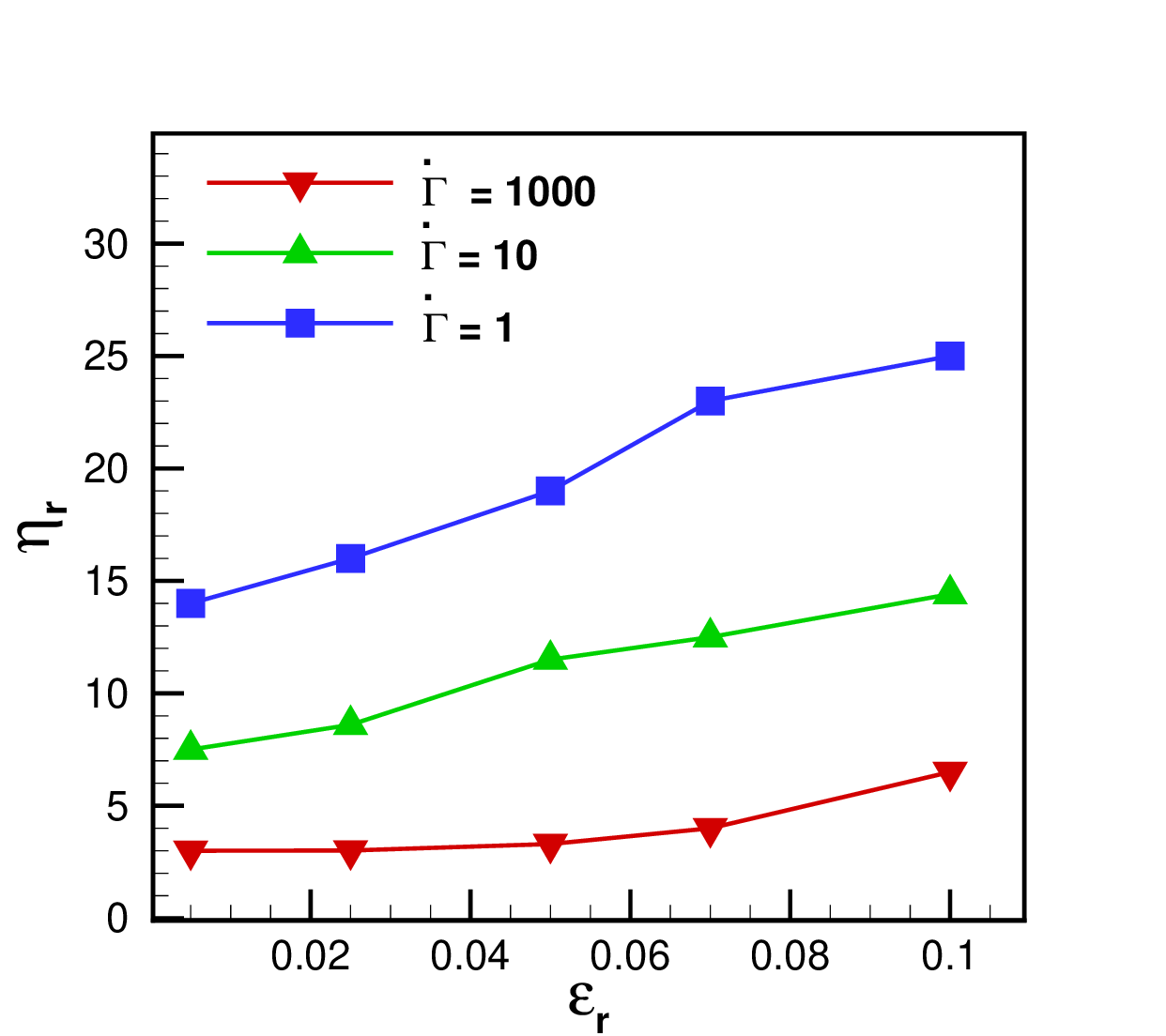}
  \caption{}
  \label{fig:roughness}
\end{subfigure}%
\begin{subfigure}{.5\textwidth}
  \centering
  \includegraphics[width=1.0\linewidth]{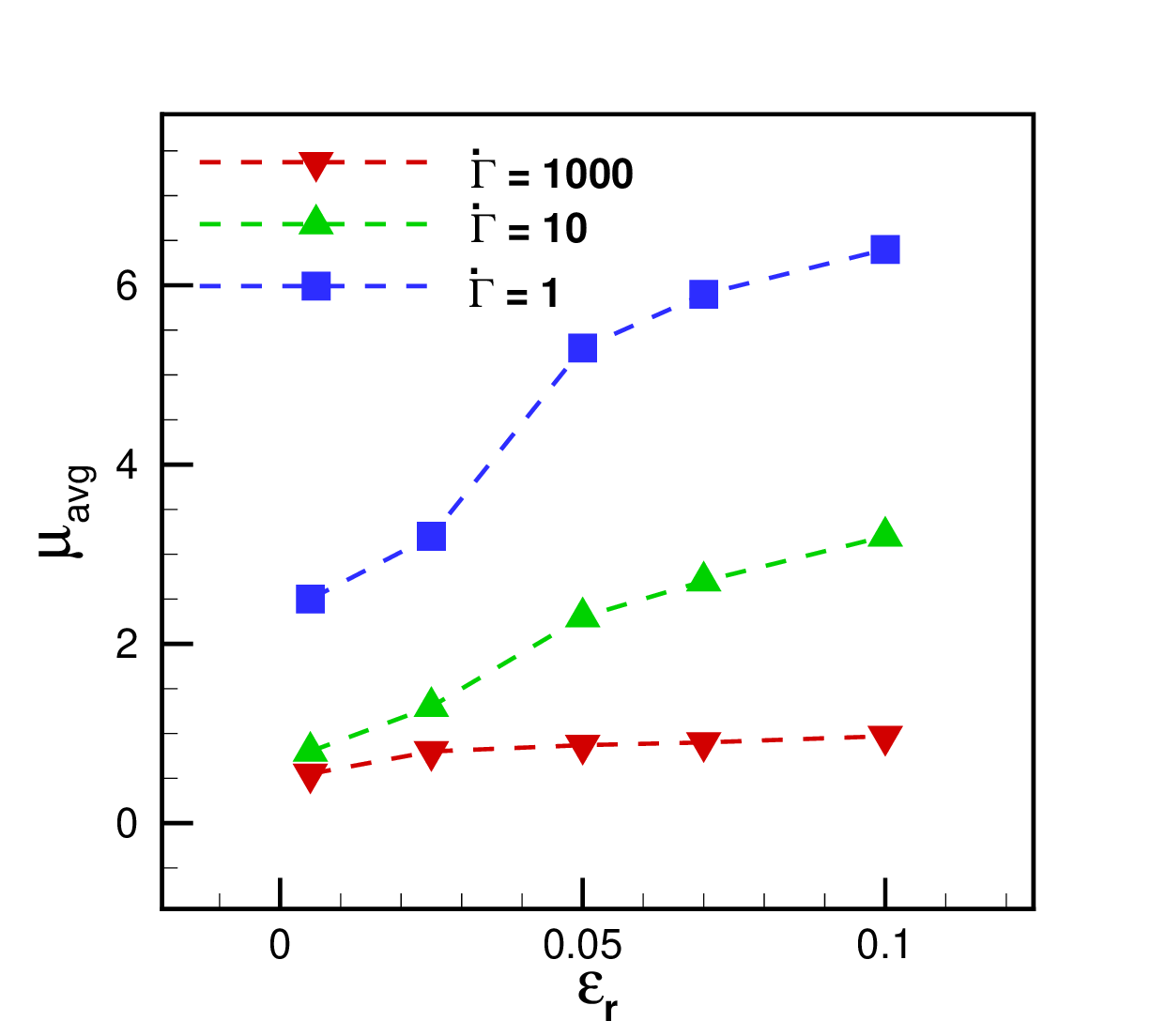}
  \caption{}
  \label{fig:mu_rough}
\end{subfigure}
\begin{subfigure}{.5\textwidth}
  \centering
  \includegraphics[width=\linewidth]{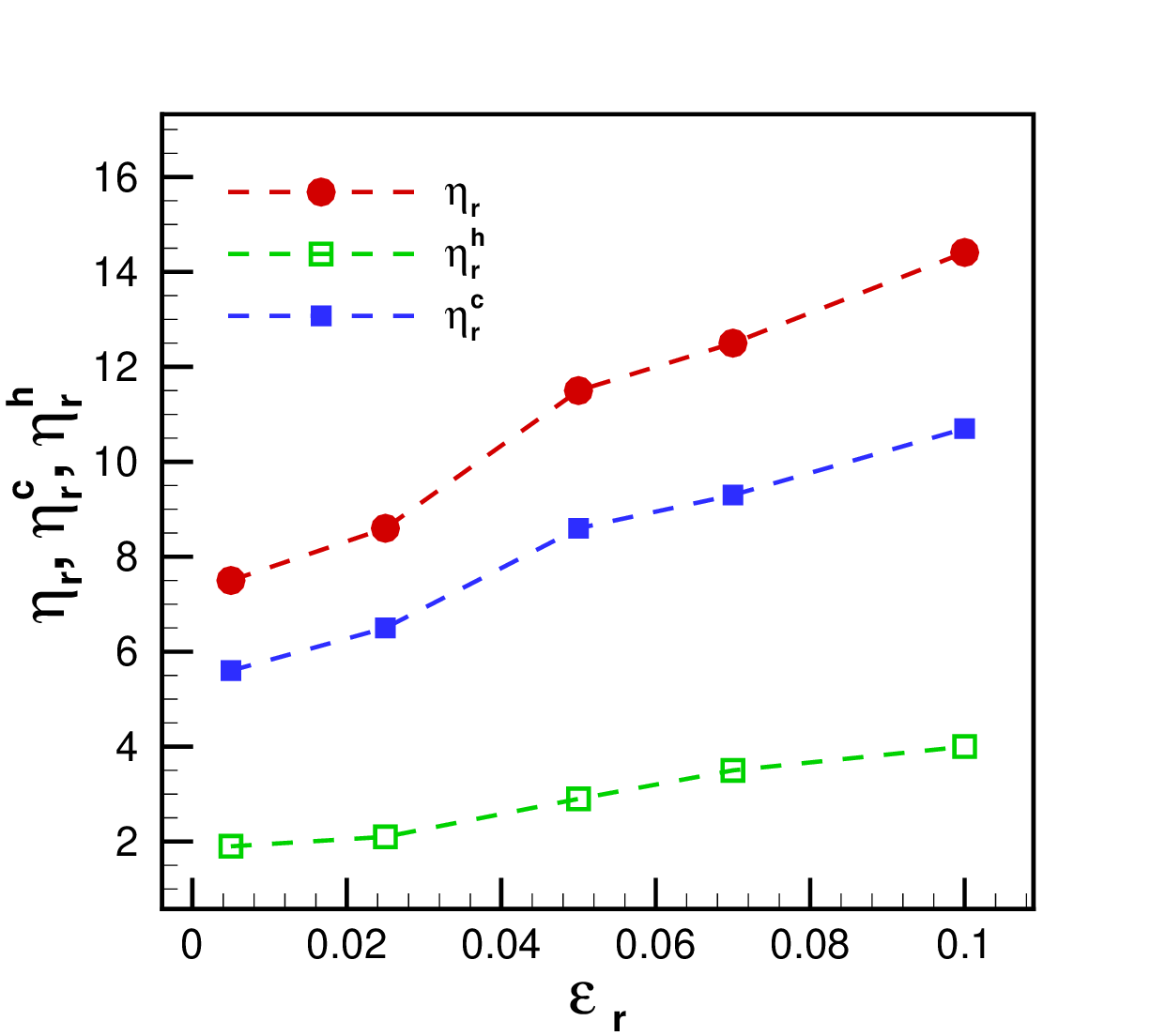}
  \caption{}
  \label{fig:stress_budget_Rough}
\end{subfigure}
\caption{(a) The relative viscosity of fiber suspension vs. the roughness height, and (b) variation in the average coefficient of friction between fibers in the suspensions with roughness at three different shear rates. For these simulations volume fraction is fixed to $\phi = 15\%$,  aspect ratio to $AR = 16$, and bending rigidity to $\tilde{B} = 0.07$. The relative viscosity increases with the increase in the surface roughness of the fibers as a consequence of increase in the $\mu_{avg}$. (c) Contribution  to the total relative viscosity from contact and hydrodynamic interactions at a fixed shear rate, $\dot{\Gamma}=10$.}
\end{figure*}

\begin{figure*}
\centering

\begin{subfigure}{.5\textwidth}
  \centering
  \includegraphics[width=1.0\linewidth]{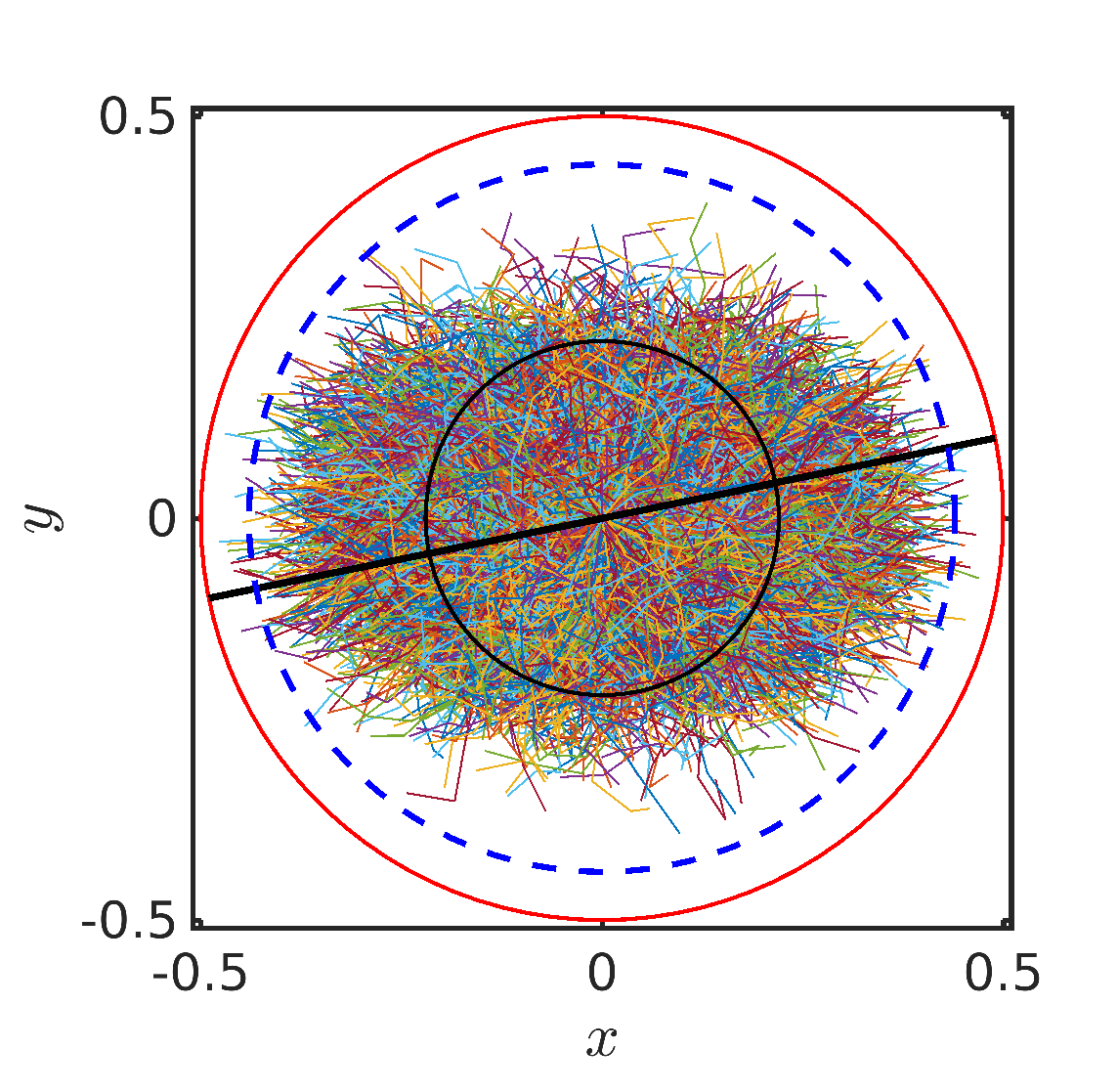}
  \caption{}
  \label{fig:smooth}
\end{subfigure}
\begin{subfigure}{.5\textwidth}
 \centering
  \includegraphics[width=0.85\linewidth]{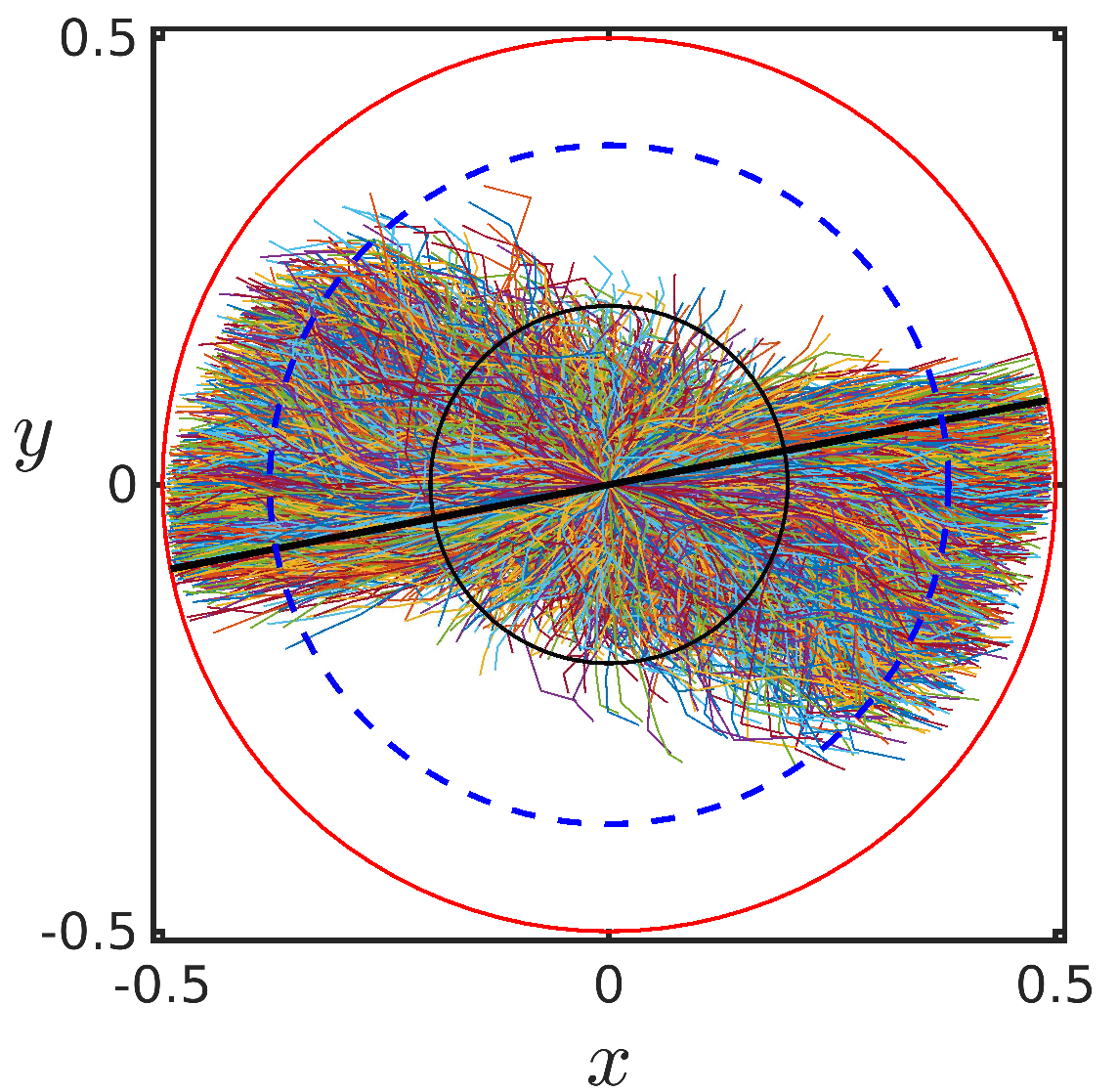}
  \caption{}
  \label{fig:rough}
  \end{subfigure}
\caption{Co-located filaments projected on the shear plane for roughness (a)  $\epsilon_r = 0.005$, and (b) $\epsilon_r = 0.05$.  The solid red line represents the circles with diameter equal to the fiber length, the blue dashed line is the circle with diameter equal to the mean end to end distance and the black circles have diameter equal to the minimum end to end distance between fibers. The solid black line shows the average orientation of the fibers with respect to the wall. All simulations were performed with dimensionless shear rate $\dot{\Gamma}=10$, volume fraction $\phi=15\%$, aspect ratio $AR = 16$, and bending rigidity  $\tilde{B}=0.07$.}
\label{fig:visu_rough}
\end{figure*}

In figure~\ref{fig:visual_3D}, we present snapshots of a thin slice in the middle of the simulation cell for suspension of fibers with roughness $\epsilon_r = 0.005$ and $\epsilon_r = 0.05$.  We observe that at high surface roughness fibers tend to agglomerate more compared to smooth fibers, increasing the overall flow resistance of the suspension. We quantify the effect of roughness by showing the increase in relative viscosity with roughness for a suspension with 15\% volume fraction at three different shear rates.
\textcolor{black}{The increase of viscosity with roughness is higher at low shear rates than at high shear rates.} This is directly linked to the steep variation of $\mu$ at low normal loads and it attaining a plateau at high normal loads (eq.~\ref{eq:frictionlaw} and figure \ref{fig:mu_normal_force}). The stress, which directly determines the normal force between the contacting fibers, is low at low shear rate values. \textcolor{black}{Hence, even a small change in the roughness size results in a large change in the friction at low shear rates as depicted by the variation of the average value of the friction coefficient, $\mu_{avg}$, displayed in figure~\ref{fig:mu_rough}.} On the contrary, at high shear rate values, $\mu$ does not change much with increasing the roughness as it attains a plateau (equation~\ref{eq:frictionlaw}). The increase in the friction as shown by the increase in $\mu_{avg}$ with roughness in figure~\ref{fig:mu_rough} is the reason behind the increase in the suspension viscosity with fiber roughness.
The consequences of the rise in the average inter-fiber friction with roughness can be understood by analyzing the different contributions to the total relative viscosity in the suspension as shown in figure~\ref{fig:stress_budget_Rough}. Here, we report the contribution of contact and hydrodynamic interactions to the total viscosity at different values of roughness $\epsilon_r$ for a fixed shear rate, $\dot{\Gamma}=10$. As the fiber roughness increases, the  contribution of contact forces to the total viscosity increases significantly, while the hydrodynamic contribution increases weakly.  
\textcolor{black}{We display the smooth and rough fibers co-located with centers at the origin in in  figure~\ref{fig:visu_rough}.  Increasing the roughness height ($\epsilon_r = 0.05$) causes larger deformation of the fibers compared to the smooth fibers ($\epsilon_r = 0.005$); increasing the hydrodynamic contribution. Moreover we note that the fibers' alignment with the flow direction does not change much with roughness.}

\begin{figure*}
\centering
\begin{subfigure}{.5\textwidth}
  \centering
  \includegraphics[width=1.0\linewidth]{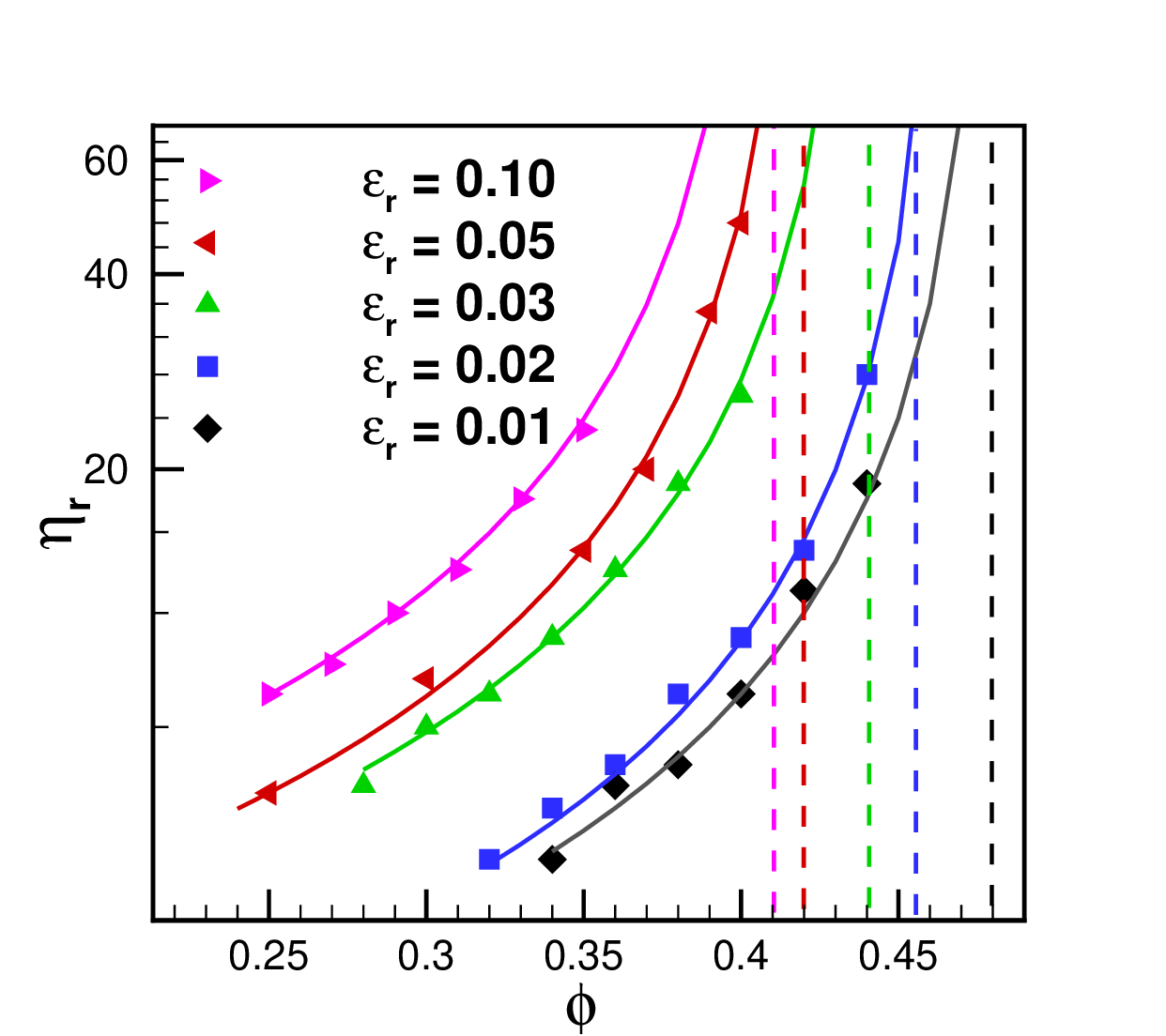}
  \caption{}
  \label{fig:roughness_jamming}
\end{subfigure}%
\begin{subfigure}{.5\textwidth}
  \centering
  \includegraphics[width=1.0\linewidth]{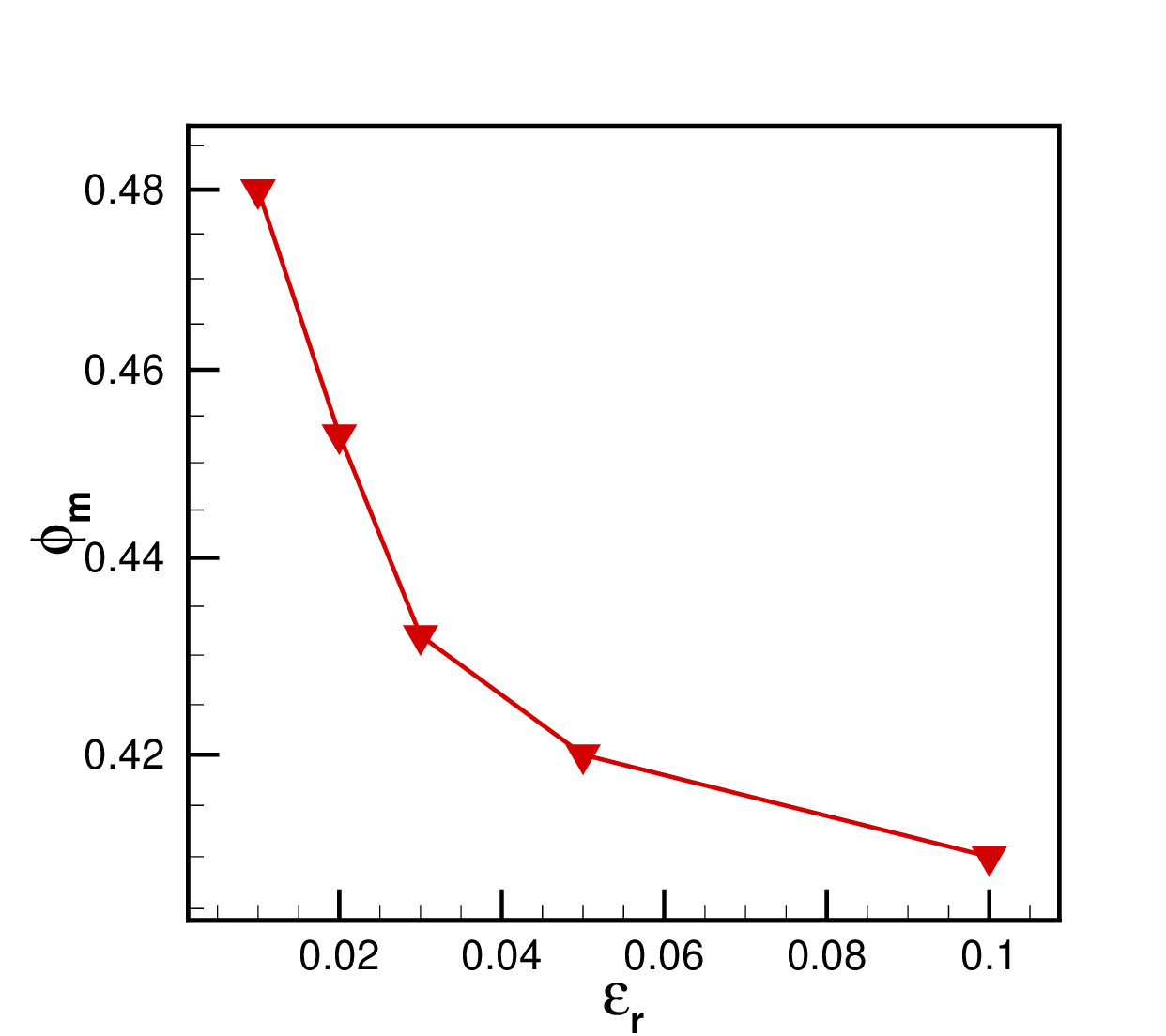}
  \caption{}
  \label{fig:roughness_maximumvol}
\end{subfigure}
\caption{(a) Relative viscosity vs volume fraction for different surface roughness values. Solid lines are the fitting curves using equation \ref{fitting}. The vertical dashed line shows the jamming volume fraction for different roughness heights. (b) The jamming volume fraction vs. roughness height. All simulations were performed for a fixed bending rigidity, $\tilde{B} = 0.07$, aspect ratio, $AR = 12$ and dimensionless shear rate, $\dot{\Gamma}=10.0$. }
\label{fig:rouhness_jamming_Full}
\end{figure*}

As increasing the fiber roughness leads to an increase in the suspension viscosity, we expect the jamming fraction to reduce with increasing the fiber roughness. The effects of varying the roughness size $\epsilon_r$ from 0.01 to 0.10 for a fixed aspect ratio $AR = 12$ is shown in figure~\ref{fig:roughness_jamming}. Again, the modified Maron-Pierce law (equation \ref{fitting}) is used to quantify the effects of roughness on the jamming fraction $\phi_m$ and the coefficient $\alpha$. The fitting parameters are presented in table \ref{tab:jamming_roughness}, whereas the solid lines in figure \ref{fig:roughness_jamming} represent the fitting  curve given by equation (\ref{fitting}). \textcolor{black}{We find that the relative viscosity diverges near the jamming transition with a scaling of $(\phi_m - \phi)^{-1}$, similarly to the cases with different aspect ratios discussed above}. These results confirm that the jamming fraction decreases with increasing the roughness size. For the smoothest case, $\epsilon_r =0.01$, we find  $\phi_m=0.48$, whereas for the roughest case, $\epsilon_r =0.1$, $\phi_m$ is as low as 0.41. The decrease in $\phi_m$ is due to an increase in the effective volume with the roughness height that results in a denser contact network \citep{more2020effect, more2020roughness}. It eventually leads to an increase in the viscosity and a decrease in the jamming volume fraction as shown in figure~\ref{fig:roughness_maximumvol}. From these findings, we can conclude that to increase the flowability of concentrated fiber suspensions, we need to break down longer fibers into smaller ones and make their surfaces smoother.

\begin{table}
  \caption{Fitting parameters for different roughness values for a fixed aspect ratio, $AR = 12$,  bending rigidity, $\tilde{B} = 0.07$ and dimensionless shear rate, $\Gamma = 10.0$ . $\beta = 1$ for all cases.}
\def~{\hphantom{1}}
  \begin{tabular}{lcc}
      $\epsilon_r$  & $\alpha$   &   $\phi_m$   \\
      \hline\\
       $0.01$   &~ $1.51$~ & ~$0.48$~ \\
       $0.02$   &~ $2.10$~ & ~$0.45$~ \\
       $0.03$  &~ $2.50$~ & ~$0.43$~ \\
       $0.05$   &~ $2.58$~ & ~$0.42$~ \\
       $0.10$   &~ $3.50$~ & ~$0.41$~ \\
       
  \end{tabular}

  \label{tab:jamming_roughness}
\end{table}

\subsection{Effects of fiber flexibility}\label{sec:flex}

\begin{figure*}
\centering
\begin{subfigure}{.5\textwidth}
 \centering
  \includegraphics[width=1.0\linewidth]{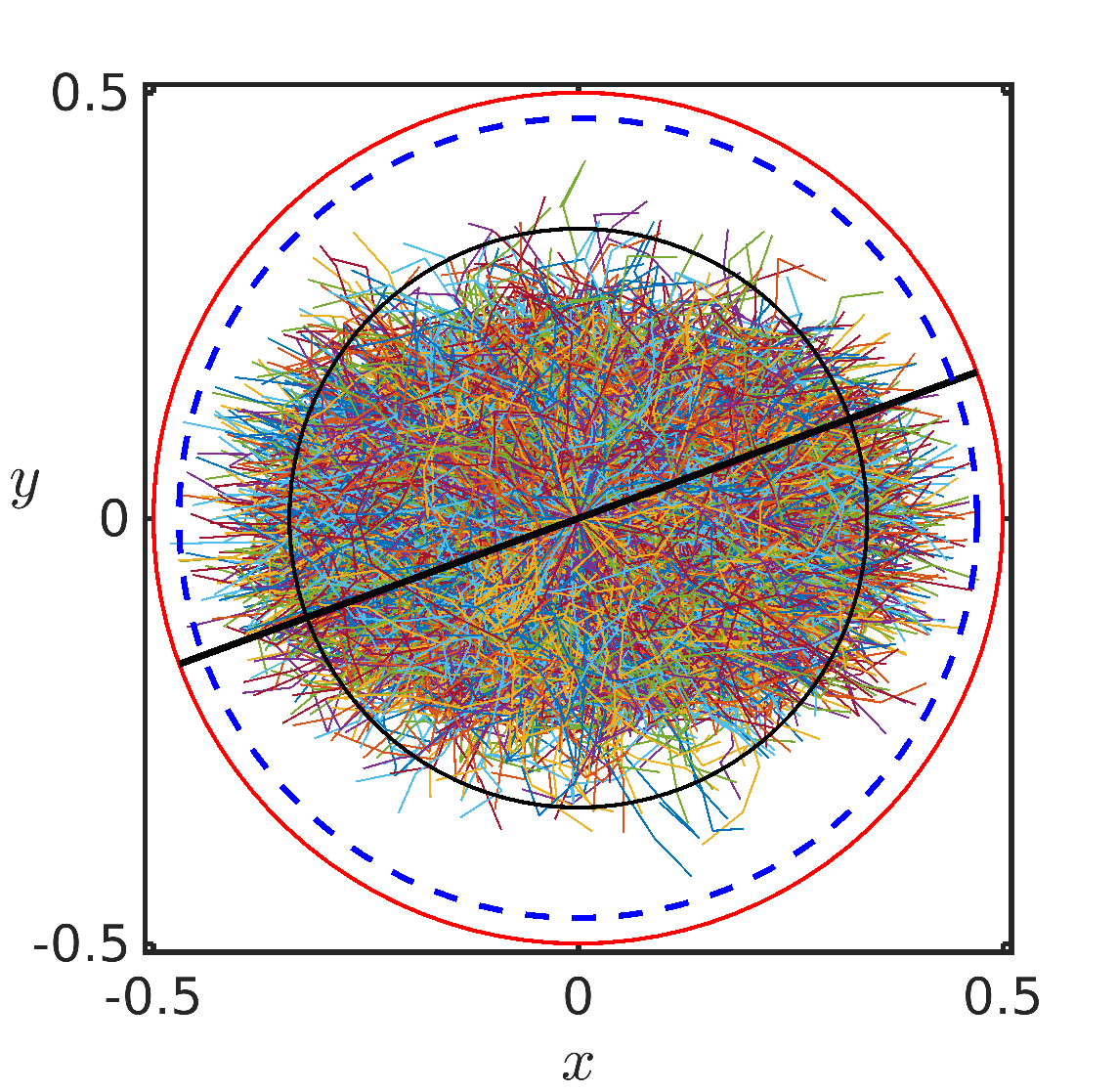}
  \caption{}
  \label{fig:rigid}
\end{subfigure}
\begin{subfigure}{.5\textwidth}
  \centering
  \includegraphics[width=1.0\linewidth]{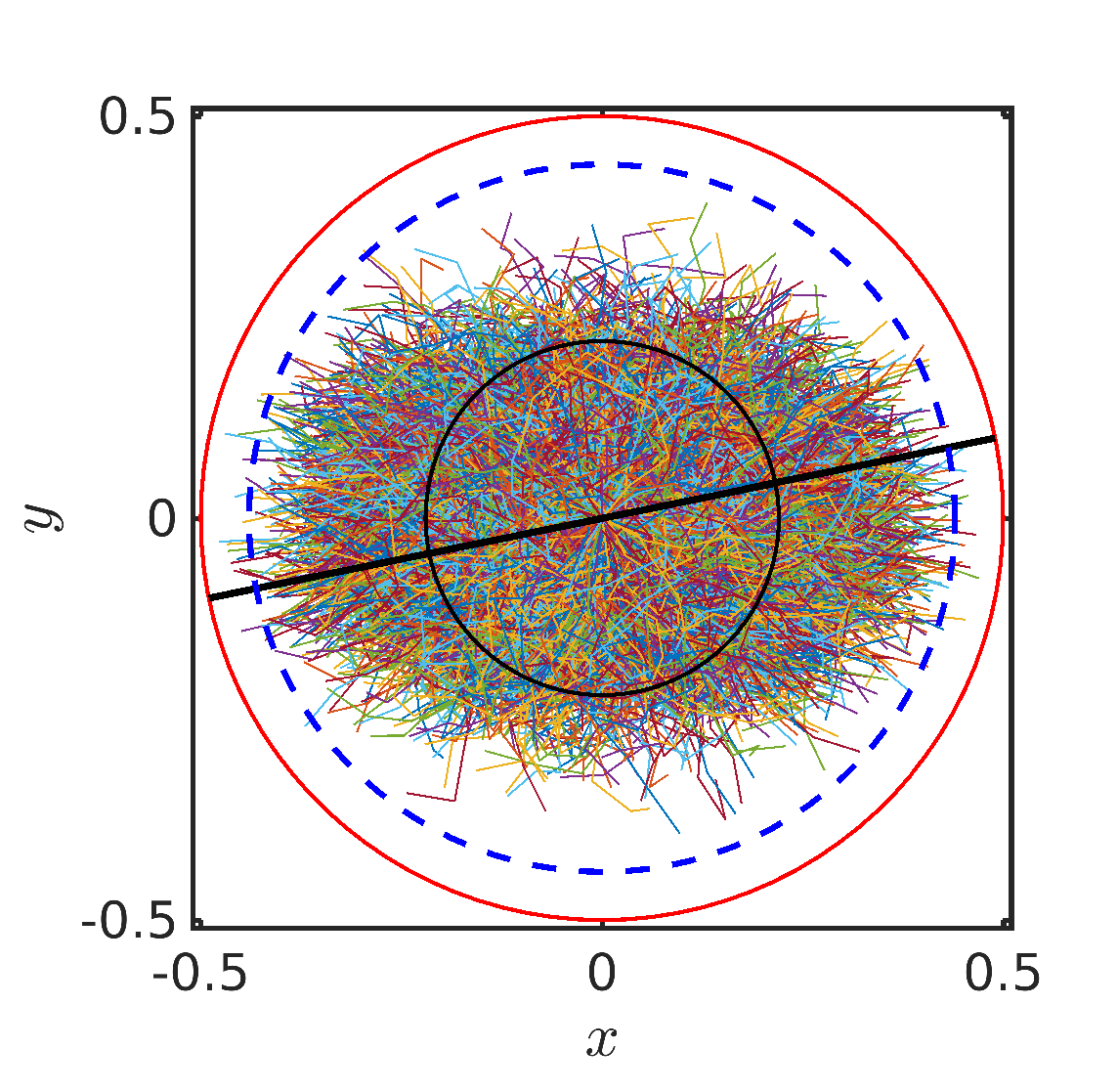}
  \caption{}
  \label{fig:flex}
\end{subfigure}
\caption{Co-located filaments projected on the shear plane for bending rigidity  (a) $\tilde{B}=0.10$, (b) $\tilde{B}=0.07$. The solid red line represents the circles with diameter equal to the fiber length, the blue dashed line is the circle with diameter equal to the mean end to end distance and the black circles have diameter equal to the minimum end to end distance between fibers. The solid black line shows the average orientation of the fibers with respect to the wall. All simulations were performed with dimensionless shear rate $\dot{\Gamma}=10$, roughness $\epsilon_r = 0.005$, volume fraction $\phi=15\%$, and aspect ratio $AR = 16$. Increasing flexibility aligns the fibers with the flow.}.
\label{fig:micro_flex}
\end{figure*}

\begin{figure*}[!t]
\centering
\begin{subfigure}{.5\textwidth}
  \centering
  \includegraphics[width=1.0\linewidth]{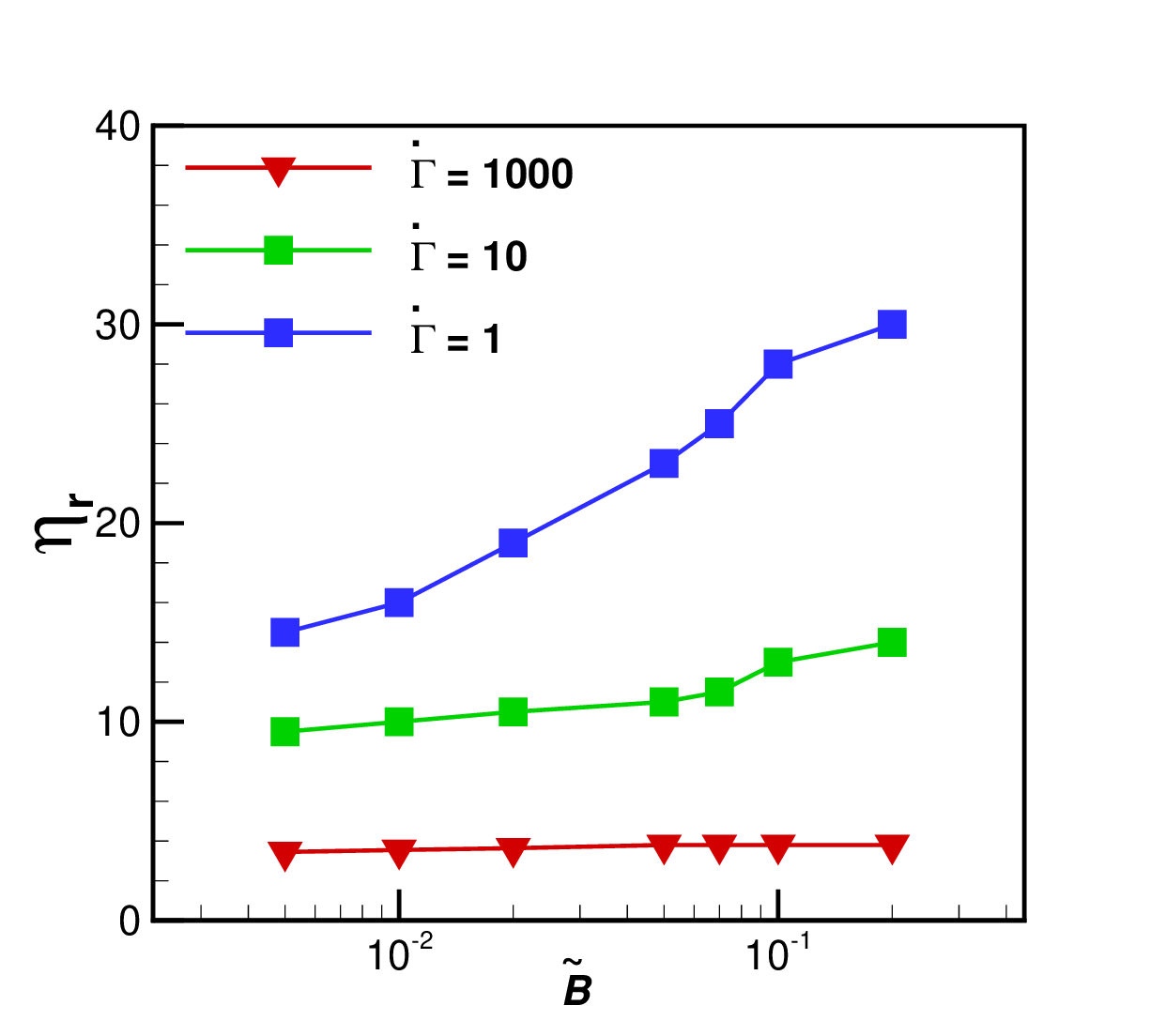}
  \caption{}
  \label{fig:Shear_bending}
\end{subfigure}%
~
\begin{subfigure}{.5\textwidth}
  \centering
  \includegraphics[width=1.0\linewidth]{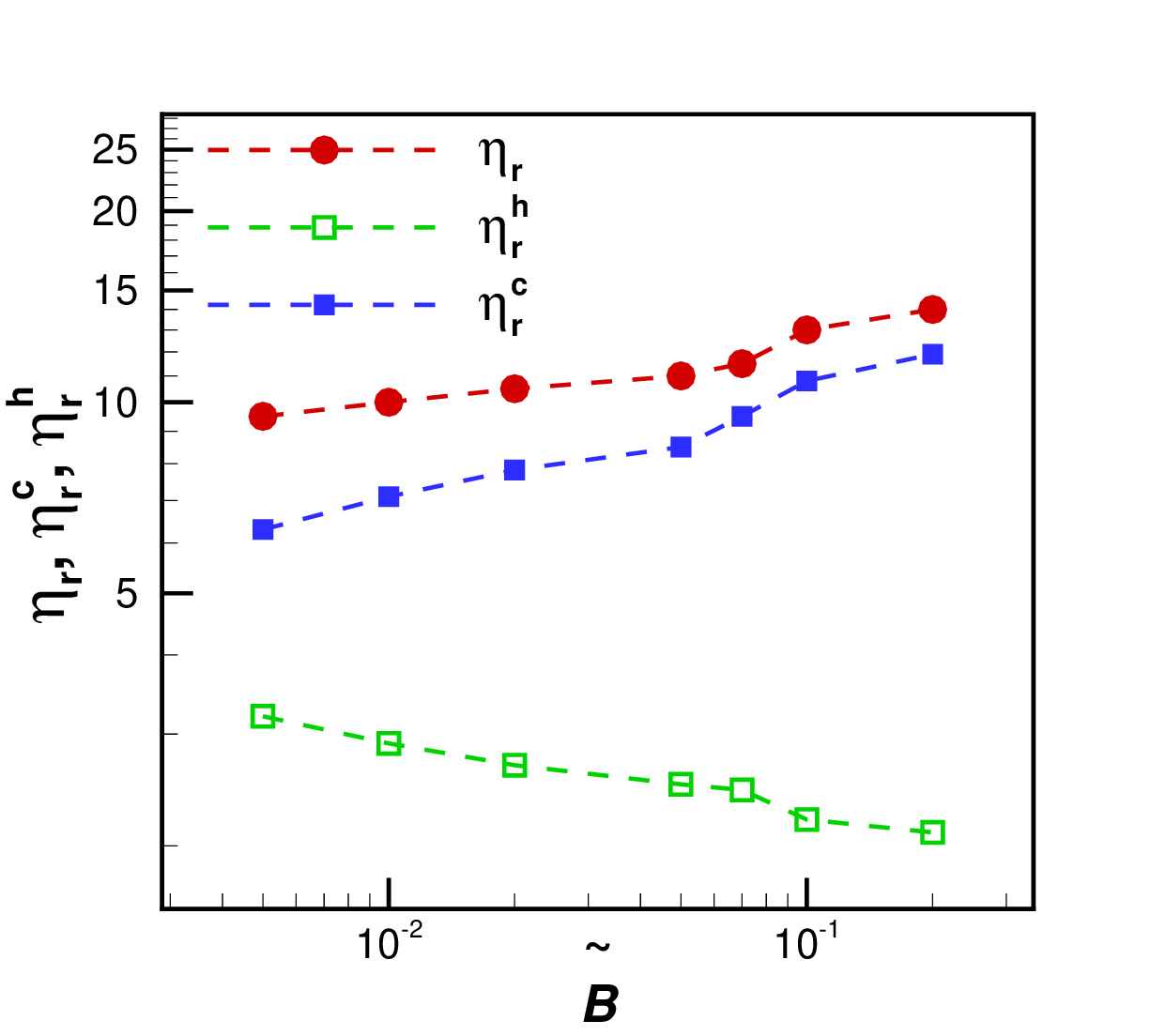}
  \caption{}
  \label{fig:Stress_Flex}
\end{subfigure}
~
\begin{subfigure}{.5\textwidth}
  \centering
  \includegraphics[width=1.0\linewidth]{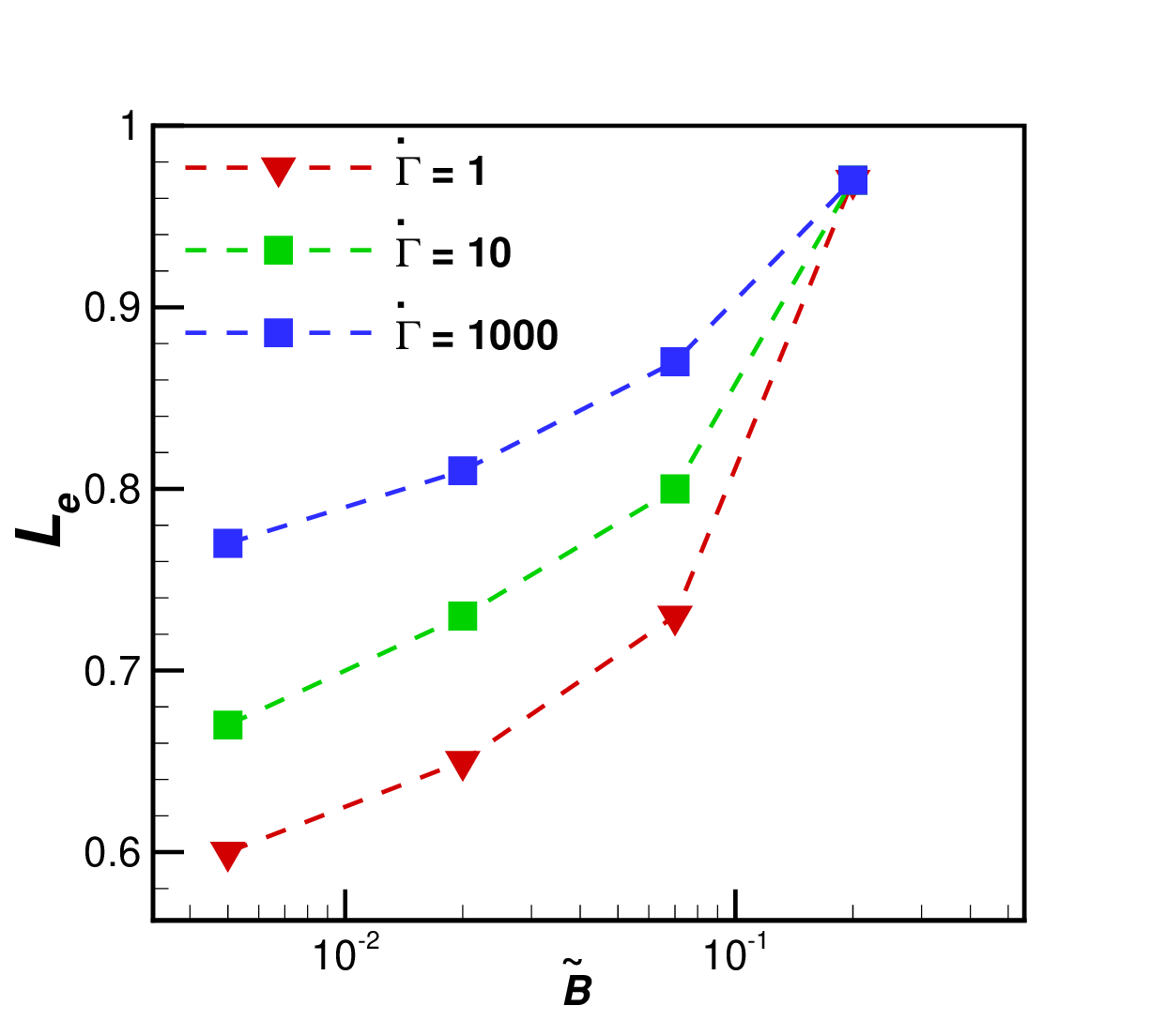}
  \caption{}
  \label{fig:end_Flex}
\end{subfigure}
\caption{(a) Relative viscosity vs. bending rigidity at three different shear rates, (b) the contributions to the total relative viscosity from contact and hydrodynamic interactions for $\dot{\Gamma}=10$, and (c) ensemble averaged end-to-end distance $L_e$ of the suspended fibers as a function of their bending rigidity for different shear rates. All the cases consider have volume fraction $\phi = 15\%$,  roughness $\epsilon_r = 0.05$,  and aspect ratio $AR = 16$.  Increase in the viscosity with the bending rigidity is consistent  with  experiments \citep{goto1986flow, sepehr2004rheological}.}
\label{fig:shear_Bending_vis}
\end{figure*}

Flexible fibers, like deformable spheres, align with the flow direction more easily than stiff fibers as shown in figure~\ref{fig:micro_flex}, thus, causing lower resistance to their flow compared to rigid fibers. As a result, flexible fiber suspensions have lower viscosity compared to rigid fiber suspensions as demonstrated in figure \ref{fig:Shear_bending}; in agreement with experiments \citep{goto1986flow,sepehr2004rheological, switzer2003rheology}. However, simulations with fibers modelled as chains of beads had predicted a rise in the suspension viscosity with an increase in the fiber flexibility \citep{wu2010numerical}, which is inconsistent with experiments \citep{goto1986flow, sepehr2004rheological}. The bending and twisting of the individual beads at their joints cause additional stresses than for continuous fibers that can only bend, and hence, result in the observed rise in suspension viscosity with flexibility. The splitting of the total relative viscosity in the hydrodynamic and the contact contributions reveals the impact of fiber flexibility on the fiber dynamics as depicted in figure~\ref{fig:Stress_Flex}. An increase in the fiber flexibility ($\tilde{B}$ decreases) increases the hydrodynamic contribution. As a result,  flexible fibers (lower $\tilde{B}$) tend to be curved instead of maintaining a straight shape as shown in figure~\ref{fig:end_Flex}, where we present time and ensemble averaged end-to-end distance $L_e$ of the suspended fibers as a function of their bending rigidity for different shear rates. A curved fiber has a larger tumbling frequency than a straight fiber, and therefore, develops larger hydrodynamic stresses \citep{tozzi2008correlation}. It is also clear that the contact contribution, which dominates the hydrodynamic contribution, increases with the fiber rigidity and is the main responsible for the higher relative viscosity at higher bending rigidity.

\subsection{Normal stress differences}\label{sec:normal}

\begin{figure*}

\begin{subfigure}{.5\textwidth}
  {\includegraphics[width=1.0\linewidth]{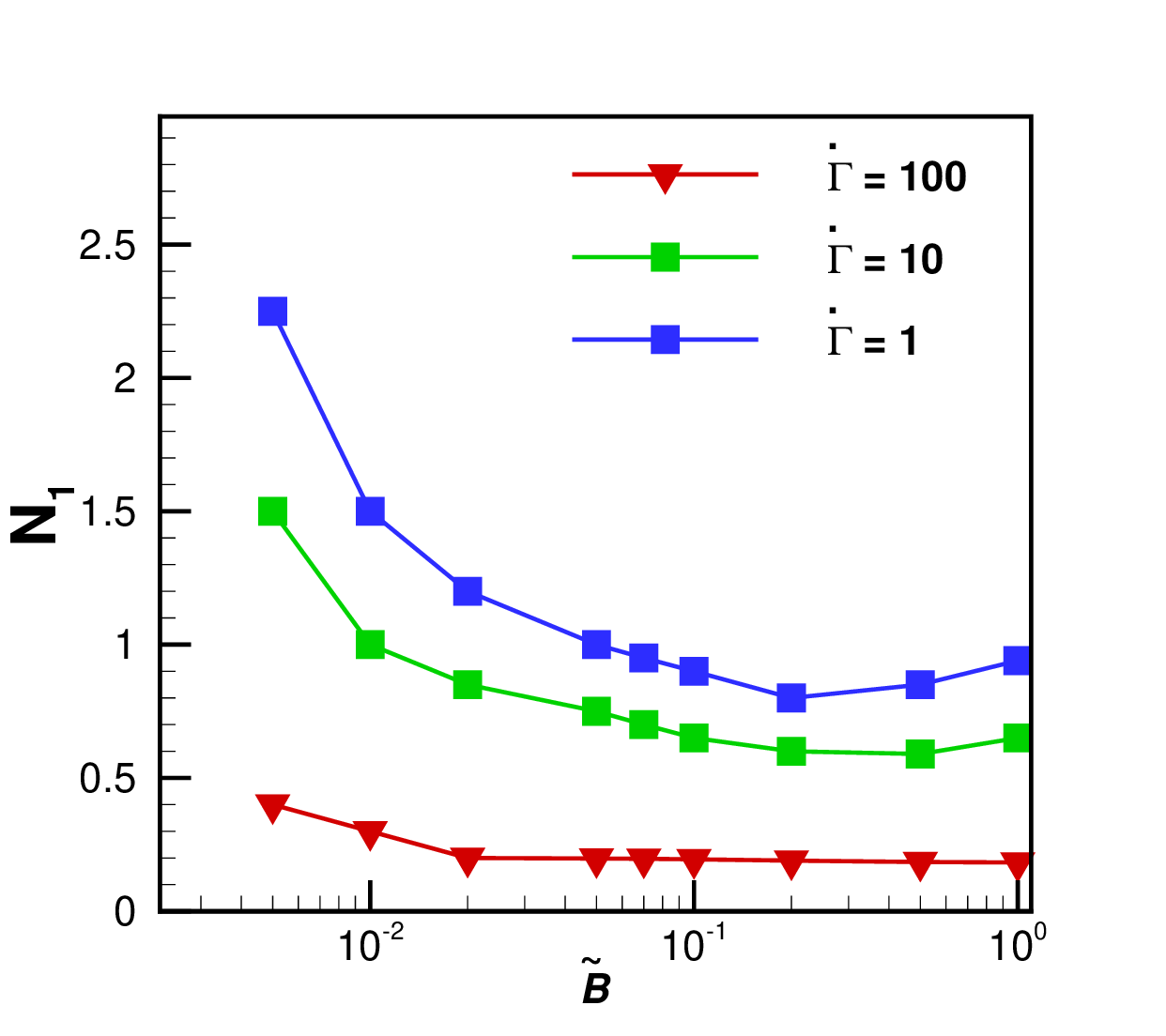}}
 \caption{}\label{fig:1st_normal_bending_shear}
\end{subfigure}%
\begin{subfigure}{.5\textwidth}
  \centering
  \includegraphics[width=1.0\linewidth]{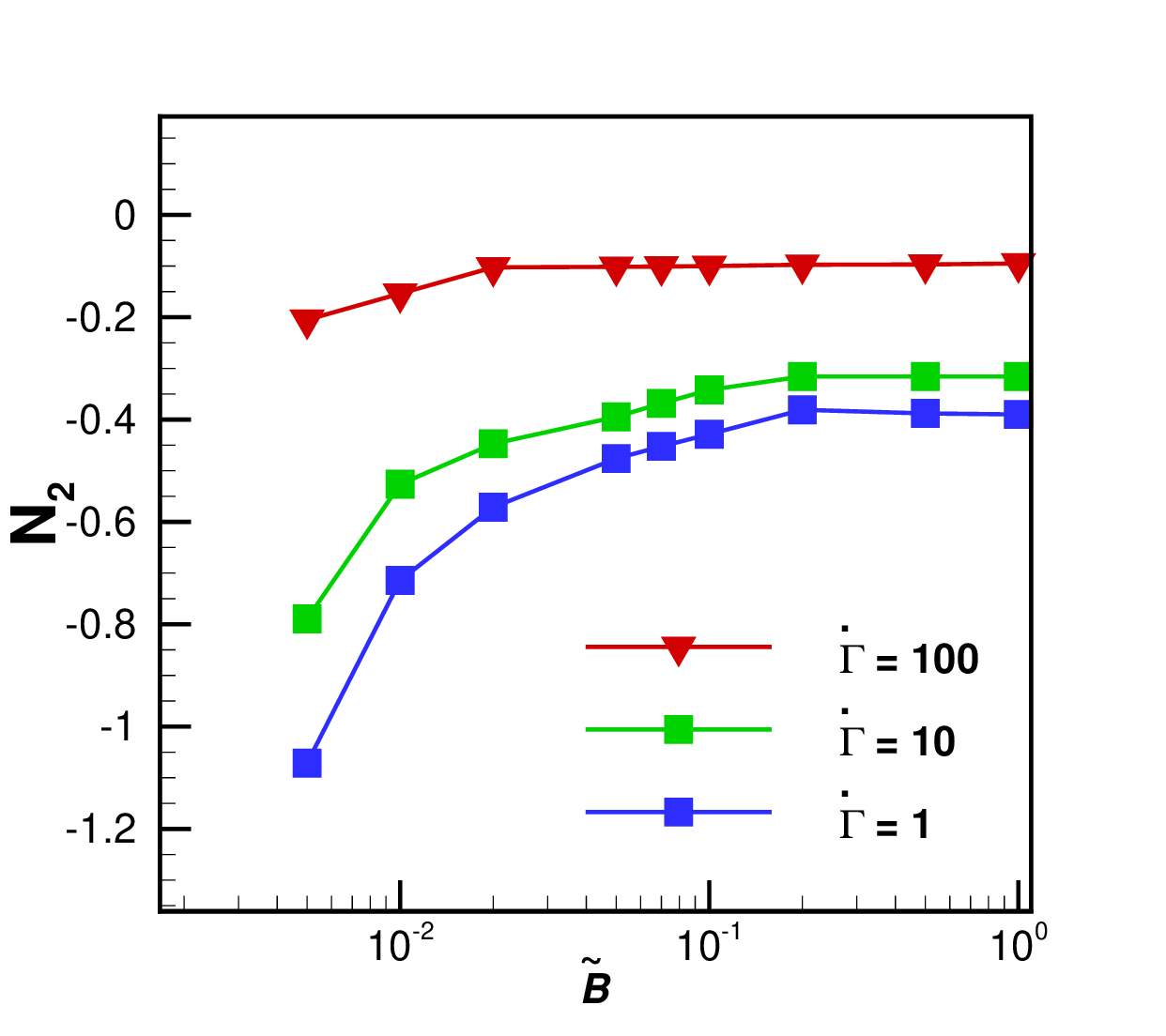}
  \caption{}
  \label{fig:N2_bending}
\end{subfigure}%
\caption{a) The dimensionless first, and b) second normal stress as a function bending rigidity for different shear rates. Here volume fraction $\phi=15\%$, roughness $\epsilon_r=0.05$, and aspect ratio $AR = 16$ are fixed. The first normal stress difference is positive and  decreases as the fibers become more rigid; in agreement with the observation of \citet{keshtkar2009rheological}. The second normal stress is negative with a decreasing magnitude with increasing fiber rigidity.}
\label{fig:N1N2}
\end{figure*}

\begin{figure}
\centering

\includegraphics[width=0.5\linewidth]{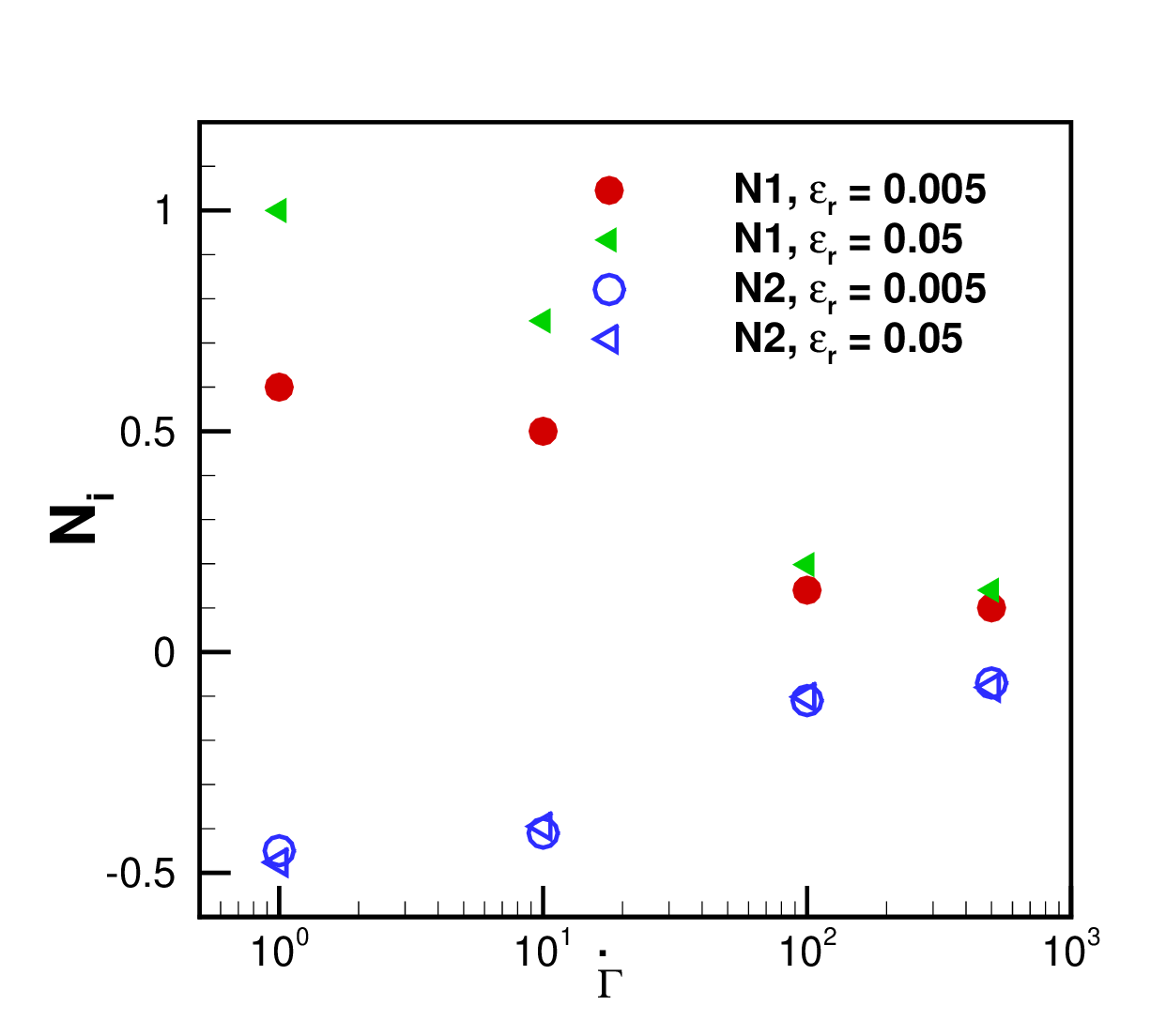}

\caption{Variation of $N_1$ and $N_2$ with shear rate for $\phi=15\%$ with a constant bending rigidity, $\tilde{B}=0.05$ ($\bigcirc$ for $\epsilon_r$ = 0.005 and $\triangleleft$ for $\epsilon_r = 0.05$). Filled symbols are for $N_1$ and hollow symbol are for $N_2$. All simulations were performed with volume fraction $\phi=15\%$ and aspect ratio $AR = 16$. This graph indicates that the surface roughness increases $N_1$ while the $N_2$ stays nearly constant.}\label{fig:N1_N2_roughness}
\end{figure}

In fiber suspensions, normal stress differences inevitably arise due to the presence of hydrodynamics and inter-particle interactions  \citep{keshtkar2009rheological}. We present the effect of fiber flexibility on the first ($N_1$) and second ($N_2$) normal stress difference in figure \ref{fig:N1N2}.  The first normal stress difference $N_1$ is positive as shown in figure~\ref{fig:1st_normal_bending_shear}, in agreement with previous studies \citep{snook2014normal,keshtkar2009rheological}. We notice that as the fiber rigidity increases, the first normal stress difference starts to decrease, in agreement with experiment reported by \citet{keshtkar2009rheological}. As the fiber rigidity increases further, we observe that there is a weak minimum for the lowest shear rate, which is also in agreement with previous reported theoretical \citep{batchelor1971stress} and computational studies \citep{wu2010numerical}.  Furthermore, the location of minimum $N_1$ shifts towards larger $\tilde{B}$ as the shear rate in increased.  Moreover, we note that to find critical bending ratio at different shear rate, we need to explore very stiff fibers like \citet{wu2010numerical}, however that is out of scope for our current study due to time step constrain.
On the other hand, the second normal stress $N_2$ is negative as shown in figure~\ref{fig:N2_bending}. Moreover, the magnitudes of  $N_2$ decrease as the fibers become more rigid,  with a more (less) pronounced decrease at lower (high) shear rates. \textcolor{black}{ Several studies with rigid fibers observed a much smaller second normal stress compared to the first normal stress which can therefore be neglected in different applications. \citep{petrich2000experimental,sepehr2004rheological}}. On the contrary, experimental and numerical study of rigid fibers by \citet{snook2014normal} indicated non-negligible second normal stress, which is approximately half of the first normal stress difference. We observe that the magnitude of $N_2$ is smaller than that of $N_1$, reported in figure \ref{fig:N1N2}. Finally, figure \ref{fig:N1_N2_roughness} presents the variation of $N_1$ and $N_2$ with the shear rate $\dot{\Gamma} $ for $15\%$ volume fraction for smooth ($\epsilon_r = 0.005$) and rough ($\epsilon_r = 0.05$) fiber suspensions. \textcolor{black}{We notice that $N_1$ increases with the roughness height, whereas $N_2$ does not change noticeably, similarly to the case of suspensions of rough spheres \citep{more2020effect}}.

\section{Conclusion}

We performed numerical simulations to investigate the rheological behavior of concentrated fiber suspensions for a range of volume fraction $\phi$,  aspect ratios $AR$, bending rigidity $\tilde{B}$, and roughness $\epsilon_r$. We modelled fibers as one-dimensional in-extensible slender bodies suspended in an incompressible Newtonian fluid. The fiber dynamics is resolved using the Euler-Bernoulli beam equation, while the immersed boundary method is utilized to resolve the fluid-fiber interactions. In addition, we modelled the fiber roughness as a hemispherical asperity on the fiber surface. As the fibers come into contact, the asperity deforms giving rise to normal and tangential contact forces and consequently friction. We used Coulomb's law of friction with a friction coefficient $\mu$ decreasing with the normal contact force as a more accurate description than a constant coefficient of friction. \textcolor{black}{Such a load-dependent friction coefficient has been measured experimentally and confirmed using a finite-element analysis \citep{brizmer2007elastic}. The numerical model presented here can quantitatively capture the experimentally observed shear thinning in fiber suspensions \citep{bounoua2016shear} as well as the variations in normal stress differences with different governing parameters.}

Our results show that the suspension relative viscosity $\eta_r$ increases with increasing volume fraction $\phi$, roughness $\epsilon_r$, and bending rigidity $\tilde{B}$. As volume fraction increases, the average number of contacting fibers $<n_c>$ increases and hence, the relative contribution from contact interactions to the bulk stress increases. This causes the observed rise in the relative viscosity. \textcolor{black}{The increase in relative viscosity with roughness is mainly attributed to the increase of the contact-force contribution to the total stress due to the rise in the average coefficient of friction, $\mu_{avg}$.} This also results in an increase in the normal stress differences $N_1$ and $N_2$. \textcolor{black}{Because of the steep variation of the friction coefficient with  the normal load (see the experimentally validated friction model in equation \ref{eq:frictionlaw})} at low normal loads (hence, low stress in the suspension), a greater increment in relative viscosity $\eta_r$ with roughness $\epsilon_r$, volume fraction $\phi$, and bending rigidity $\tilde{B}$ at lower shear rates is observed. On the other hand, the  friction coefficient $\mu$ attains a plateau at high normal loads (hence, high stress), which also results in a flattening of the relative viscosity $\eta_r$ with roughness, volume fraction, and bending rigidity at higher shear rates.

\textcolor{black}{We have presented the macroscopic flow properties of the suspensions and discussed the stress budget as well as fiber deformation for varying the fiber rigidity and roughness. Our results show that  more rigid fibers resist deformation and do not align with the flow. As a result, they create less ordered structures which increase the resistance to the shear flow, consequently increasing the suspension viscosity. Moreover, we find that fibers deform more when their surface roughness increases, an effect similar to that observed when increasing the fiber flexibility;  however, rough fibers do not align with the flow as flexible fibers. As a result, the relative viscosity increases with increasing roughness, but it decreases when increasing the fiber flexibility. So, even though the mesoscopic effect (fiber deformation and alignment) of these microscopic parameters (flexibility and roughness) is the same, they results in a different macroscopic flow behavior. Furthermore, our results for the normal stress differences showed that the first normal stress difference is positive and the second normal stress difference is negative, which is consistent with previous studies \citep{keshtkar2009rheological,snook2014normal}. Flexible fiber suspensions have a larger magnitude of the first  and second normal stress differences, also in agreement with available experiments. \cite{keshtkar2009rheological}.}

We also explored the divergence of the relative viscosity as the suspension volume fraction approaches the maximum flowable limit, i.e, the jamming fraction $\phi_m$. We presented a modified Maron-Pierce law to quantify the effects of the governing parameters on $\phi_m$. In particular, we show that increasing $AR$ and $\epsilon_r$ decrease $\phi_m$. Re-scaling $\phi$ by $\phi_m$ collapses the $\eta_r$ data for varying $AR$ on a single curve denoting that changing the fiber aspect ratio principally affects the maximum volume fraction if the roughness is fixed. The decrease in the jamming volume fraction with aspect ratio is attributed to the increase in the number of contacts and the contact forces, which in turn increases the stress in the suspension at the same shear rate. We also quantified the jamming fraction for different roughness heights. As the roughness height increases, the suspension jams at a lower volume fraction. With an increase in the roughness height, the effective volume fraction increases, resulting in a denser contact network. As a result the average coefficient of friction increases, which in turn reduces the jamming volume fraction. We found the viscosity to diverge as $(\phi_m - \phi)^{-1}$, in contrast with spherical suspensions, where viscosity diverges as $(\phi_m - \phi)^{-2}$.

Our results demonstrate the importance of accurately modelling the inter-fiber interactions to capture the experimentally observed shear rate-dependent rheological behavior of fiber suspensions. We showed that the fiber tribology governs the suspension flow behavior. Since it is challenging to modify the coefficient of friction, we can modify the fiber surface properties that influence the friction (e.g., roughness) to tune the suspension flow properties. Additionally, we found that to have higher solid concentrations, desired in industrial applications, we should breakdown the fibers into smaller ones, pre-treat them to be more flexible, and to make the surface smoother. Finally, the mono-asperity contact model implemented here could be refined by taking into account a finite number of asperities, possibly growing with the load. \textcolor{black}{In addition, it will be an interesting study to  find how the model to predict anisotropic diffusivity changes from previous study \cite{salahuddin2012numerical} when the effect of friction is considered for numerical simulations of dense fiber suspension.} More generally, due to the complexity of the friction forces at the microscopic scale and the number of parameters that are potentially relevant in the surface physical chemistry, the next step would be to quantitatively determine the frictional law from colloidal probe AFM measurements.

\textbf{ACKNOWLEDGEMENT}

AMA would like to acknowledge financial support from the Department of Energy via grants EE0008256 and EE0008910. This work used the Extreme Science and Engineering Discovery Environment (XSEDE) \citep{towns2014xsede}, which is supported by the National Science Foundation grant number ACI-1548562 through allocations TG-CTSI190041. L.B. acknowledges financial support from the Swedish Research Council (VR) and the INTERFACE research environment (Grant No. VR 2016-06119). 

\bibliographystyle{jfm}

\bibliography{paper}

\end{document}